\documentclass[journal,twoside,web]{ieeecolor}
\usepackage{cite}
\usepackage{amsmath,amssymb,amsfonts}
\usepackage{algorithmic}
\usepackage{booktabs}
\usepackage{pgfplots, tikz}
\pgfplotsset{compat=1.18}
\pgfplotsset{every axis/.append style={very thick}}
\usepackage{graphicx}
\usepackage{textcomp}
\usepackage{xcolor}
\definecolor{sectioncolor}{rgb}{0,0.541,0.855}
\definecolor{subsectioncolor}{rgb}{0,0.541,0.855}
\def\logoname{LOGO-tmi-web}
\def\journalname{IEEE Transactions on Medical Imaging}
\def\BibTeX{{\rm B\kern-.05em{\sc i\kern-.025em b}\kern-.08em
    T\kern-.1667em\lower.7ex\hbox{E}\kern-.125emX}}
\begin{document}
\title{Hybrid Multihead Attentive Unet-3D for Brain Tumor Segmentation}
\author{Muhammad Ansab Butt, Absaar Ul Jabbar
\thanks{Muhammad Ansab Butt is with the
University of Engineering and Technology, Lahore, Pakistan (e-mail:engransab4@gmail.com). }
\thanks{Absaar Ul Jabbar is with the School of Interdisciplinary Engineering and Sciences, National University of Sciences and Technology, Islamabad, Pakistan (e-mail: absaar@sines.nust.edu.pk, absaar468@gmail.com).}
}

\maketitle

\begin{abstract}
Brain tumor segmentation is a critical task in medical image analysis, aiding in the diagnosis and treatment planning of brain tumor patients. The importance of automated and accurate brain tumor segmentation cannot be overstated. It enables medical professionals to precisely delineate tumor regions, assess tumor growth or regression, and plan targeted treatments. Various deep learning-based techniques proposed in the literature have made significant progress in this field, however, they still face limitations in terms of accuracy due to the complex and variable nature of brain tumor morphology. In this research paper, we propose a novel Hybrid Multihead Attentive U-Net architecture, to address the challenges in accurate brain tumor segmentation, and to capture complex spatial relationships and subtle tumor boundaries. The U-Net architecture has proven effective in capturing contextual information and feature representations, while attention mechanisms enhance the model's ability to focus on informative regions and refine the segmentation boundaries. By integrating these two components, our proposed architecture improves accuracy in brain tumor segmentation. We test our proposed model on the BraTS 2020 benchmark dataset and compare its performance with the state-of-the-art well-known SegNet, FCN-8s, and Dense121 U-Net architectures. The results show that our proposed model outperforms the others in terms of the evaluated performance metrics.
\end{abstract}
\section{Introduction}
\label{introduction}
Brain tumor segmentation plays a vital role in the diagnosis, treatment planning, and monitoring of brain tumors. Accurate and automated segmentation of tumor regions from medical images assists in quantitative analysis, clinical decision-making, and improving patient outcomes. In recent years, deep learning-based approaches have revolutionized medical image segmentation, providing robust and precise tumor delineation. Among these approaches, the 3D U-Net \cite{7404017} \cite{chang2018brain} \cite{ballestar2020mri} \cite{ahmad2021context} \cite{baid2020novel} architecture has demonstrated remarkable success due to its ability to capture both local and global contextual information. Furthermore, the incorporation of multihead attention mechanisms enhances the model's ability to focus on relevant features, leading to improved segmentation accuracy.

Multihead attention (MHA) is a key component in modern neural network architectures, particularly in the field of Natural Language Processing (NLP). It enables neural networks to focus on different parts of the input sequence simultaneously and captures multiple perspectives or representations by splitting the input into multiple ”heads” and applying attention mechanisms independently. MHA enhances the modeling of long-range dependencies, improves representation learning by capturing diverse information, enables parallelization for efficient computation, and enhances robustness to noise and variations.
While the effectiveness and versatility of MHA have been
successfully demonstrated in various NLP applications, such as machine translation, text summarization, sentiment analysis [21], question answering [35], and language generation, its exploration in the context of segmentation tasks has been relatively limited until recent times.

\subsection{Motivation}
The incorporation of multihead attention mechanisms in segmentation algorithms can enhance their ability to focus on relevant features, leading to improved segmentation accuracy. Here, we introduce a novel approach that combines the power of MHA with the widely used  3D U-Net architecture for accurate brain tumor segmentation. By leveraging MHA's ability to capture diverse and fine-grained information, we aim to enhance the segmentation performance and address the challenges associated with brain tumor analysis, such as handling diverse tumor types, sizes, and imaging modalities. Our proposed approach holds great promise in improving the accuracy and reliability of brain tumor segmentation, paving the way for more advanced and effective medical imaging applications. By achieving accurate tumor segmentation, we aim to provide clinicians with a reliable tool for tumor characterization, treatment planning, and monitoring of tumor progression.

\subsection{Objectives}
The main objectives of this research study are:

\begin{itemize}
    \item To develop a 3D U-Net model with a multihead attention mechanism for accurate and automated brain tumor segmentation.
    \item To evaluate the performance of the proposed model on BraTS 2020 brain tumor dataset, considering different tumor types and imaging modalities.
    \item To investigate the impact of different hyperparameter values on the performance of the proposed model.
    \item To compare the performance of the proposed model with state-of-the-art segmentation methods such as SegNet, FCN-8s, and Dense121 U-Net architectures.
    \item To analyze the interpretability and generalizability of the proposed model by visualizing and assessing its segmentation results.
\end{itemize}

\subsection{Contributions}
\begin{itemize}
    \item A novel 3D U-Net architecture with multihead attention tailored specifically for brain tumor segmentation, enabling accurate tumor delineation.
    \item Extensive experiments and evaluations on benchmark brain tumor dataset BraTS 2020 demonstrate the proposed model's effectiveness and robustness.
    \item Comprehensive analysis of various factors influencing the model's performance, providing insights into optimizing the segmentation process.
    \item Performance comparison of the proposed model with the state-of-the-art segmentation methods.
    \item In-depth visualization and interpretation of the segmentation results, aiding clinicians in understanding and validating the model's performance.
    
\end{itemize}

\subsection{Practical Applications}

The accurate segmentation of brain tumors using advanced deep learning models, such as the 3D U-Net architecture with multihead attention, holds significant practical implications in various fields, including brain telemetry and brain transplantation.

In the context of brain telemetry, the ability to accurately segment brain tumors is crucial for understanding the spatial extent and location of the tumor. This information is vital for the placement and positioning of brain telemetry devices, such as Neuralink's brain chips \cite{schlageter1999microvessel}. Accurate tumor segmentation allows for precise targeting of specific regions of interest, enhancing the effectiveness of brain telemetry for monitoring brain activity and collecting data. Additionally, the segmentation of brain tumors enables the identification of critical structures and functional areas adjacent to the tumor, ensuring their preservation during the implantation of brain chips.

From the perspective of tumor surgery, accurate tumor segmentation plays a crucial role in surgical planning. Surgeons need to visualize and understand the tumor's extent and location to perform precise and successful neurosurgical procedures. Accurate segmentation helps in identifying tumor boundaries and differentiating between tumor sub-regions, such as the core, active tumor, and edema. This information guides surgeons in determining the optimal resection margins and ensuring complete removal of the tumor while preserving healthy brain tissue. Furthermore, post-surgery monitoring of tumor progression and treatment response heavily relies on accurate and reliable segmentation of tumor regions.

Therefore, the application of advanced deep learning techniques, such as the 3D U-Net architecture with multihead attention, in brain tumor segmentation can contribute to the success and advancement of practical applications such as brain telemetry and neurosurgery. These techniques can enable precise and accurate delineation of tumor regions, facilitating optimal surgical planning, patient-specific treatment strategies, and improved post-operative monitoring.

\section{Literature Review}
Brain tumor segmentation poses significant challenges in medical image analysis. The task gets complicated due to tumor heterogeneity, where tumors exhibit diverse characteristics in shape, size, texture, and intensity. To address these challenges a lot of research is being carried out to develop more accurate and efficient brain segmentation methods. The MICCAI Brain Tumor Segmentation (BraTS) Challenge has introduced the Multimodal Brain Tumor Image Segmentation Benchmark datasets \cite{menze2014multimodal} \cite{zhao2015deep}, to facilitate the evaluation of segmentation methods.

\subsection{Traditional Approaches for Brain Tumor Segmentation}

Traditional image segmentation approaches have employed techniques like thresholding, region growing, and mathematical morphology \cite{7002427} \cite{5639536}. However, these techniques suffer from several limitations when applied to brain tumor segmentation. 

The thresholding technique, for instance, relies heavily on intensity values to distinguish tumor regions from the surrounding healthy tissue. However, brain tumor images often exhibit significant intensity variations due to factors such as image artifacts, diverse tumor appearances, and heterogeneous tissue characteristics, making it challenging to set appropriate thresholds. 

Similarly, region growing and mathematical morphology methods often focus on local image information and may not effectively capture the spatial context of the tumor. This limitation can result in fragmented segmentations or the inclusion of non-tumor regions within the segmented area.

Moreover, these segmentation methods are sensitive to noise and artifacts, leading to false positive or false negative tumor detections. Brain tumor images can suffer from inherent noise and imaging artifacts, such as partial volume effects, and motion artifacts, which can affect the accuracy of these methods. 

Furthermore, many traditional methods require manual parameter tuning to adapt to different image characteristics and tumor types. This process can be time-consuming and subjective, as the optimal parameters may vary depending on the specific dataset and user expertise. Brain tumor diagnosis often involves the integration of multiple imaging modalities, such as MRI, CT, and PET scans. Traditional segmentation methods struggle to handle the inherent challenges of multimodal imaging, such as intensity variations, registration issues, and different contrasts.

To overcome the limitations, more advanced and sophisticated techniques, such as deep learning-based approaches, have emerged in recent years, offering improved accuracy, robustness, and adaptability in brain tumor segmentation.

\subsection{Deep Learning for Brain Tumor Segmentation}

The emergence of deep learning techniques has revolutionized medical image segmentation, including brain tumor segmentation. Convolutional Neural Networks (CNNs) have demonstrated remarkable success in capturing complex and hierarchical image features, leading to improved segmentation accuracy.

Many researchers have used a CNN-based approach specifically for brain tumor segmentation using MRI images \cite{bernal2019deep} \cite{zhang2015deep} \cite{pedada2023novel}.  They employed a 2D CNN with a patch-based approach to handle large images. By dividing the input images into smaller patches, the CNN could effectively capture local information and learn discriminative features. The patch-based approach also helped mitigate memory constraints associated with processing large 3D medical images. \cite{zhu2023brain} proposed a technique based on the fusion of deep semantics and edge information in multimodal MRI.

However, the 2D CNN approach lacks full utilization of 3D information and spatial context, limiting its accuracy in capturing complex tumor variations and shapes. It also exhibits increased computational complexity compared to 3D approaches, requiring longer processing times.

\cite{kamnitsas2017efficient} introduced an efficient multi-scale 3D CNN with fully connected CRF for accurate brain lesion segmentation. Similarly, \cite{hu2019brain} introduced brain tumor segmentation using multi-cascaded convolutional neural networks and conditional random field.    In another study \cite{zhou2020afpnet},  a 3D fully connected convolutional neural network with an atrous-convolution feature pyramid was introduced for brain tumor segmentation via MRI images. \cite{xue2018segan} introduced an Adversarial network with multi-scale 1-1 loss for medical image
segmentation.

The above-mentioned models incorporated multi-scale processing to capture both local and global contextual information, enabling more accurate segmentation. The fully connected CRF further refined the segmentation results by considering spatial dependencies between neighboring voxels. This approach significantly improved brain lesion segmentation accuracy compared to traditional methods.

These studies showcased the potential of CNNs in automating the brain tumor segmentation process and achieving higher accuracy compared to traditional methods. The ability of CNNs to learn discriminative features directly from the data, combined with their ability to capture both local and global information, has made them a powerful tool in the field of medical image analysis. By leveraging the large amounts of labeled data available in brain tumor datasets, CNN-based approaches have been able to learn intricate patterns and variations, enabling robust and accurate segmentation.

However, despite the success of CNNs, challenges remain in handling class imbalance, limited training data, and addressing the interpretability of the learned features. Researchers continue to explore novel architectures, data augmentation techniques, and regularization strategies to overcome these challenges and further improve the performance of deep learning models for brain tumor segmentation.

\subsection{The U-Net Architecture for Biomedical Image Segmentation}

The U-Net architecture, proposed by \cite{ronneberger2015u}, has gained widespread adoption in biomedical image segmentation tasks, including brain tumor
segmentation\cite{futrega2021optimized}  \cite{kermi2019deep}, segmentation of skin lesions \cite{liu2019skin} etc. The U-Net architecture is renowned for capturing local and global contextual information, enabling precise and accurate segmentation.

The U-Net architecture consists of a contracting path and an expanding path, forming a U-shaped network structure. The contracting path, also known as the encoder, captures contextual information by progressively reducing the spatial dimensions of the input image. It consists of a series of convolutional and pooling layers that learn hierarchical representations of the input data. Each convolutional layer is typically followed by batch normalization and activation functions, such as ReLU, to introduce non-linearity and enhance feature extraction.

To enable precise localization, skip connections are introduced in the U-Net architecture \cite{mubashar2022r2u++}. These connections establish direct connections between the corresponding encoding and decoding layers. By merging the feature maps from the contracting path with the upsampled feature maps from the expanding path, the model can fuse both low-level and high-level features. This mechanism enables the U-Net architecture to capture fine-grained details while preserving important contextual information, leading to improved segmentation accuracy.

Various studies have presented an automated design methodology for U-Net models in biomedical image segmentation \cite{noori2019attention}\cite{aboelenein2020httu}. They introduced an optimization framework that automates the process of selecting the model architecture, including the number of layers and filters. This methodology streamlines the design process and ensures the effectiveness and flexibility of U-Net models in various biomedical image segmentation tasks.

The U-Net architecture has demonstrated remarkable performance in various biomedical segmentation tasks, including brain tumor segmentation. Its ability to capture both local and global information, coupled with the fusion of low-level and high-level features, has made it a popular choice among researchers. However, further advancements in U-Net-based models continue to be explored, including the integration of attention mechanisms and additional architectural modifications, to further enhance its segmentation accuracy and robustness.

\subsection{Transformers in Brain Tumor Segmentation}

Transformers, originally introduced by \cite{vaswani2017attention} for natural language processing tasks, have gained significant attention in various computer vision applications, including medical image segmentation. Transformers are based on the self-attention mechanism, allowing them to capture global dependencies and long-range interactions in the input data. In recent years, researchers have explored the application of Transformers in brain tumor segmentation, leveraging their ability to model complex relationships within medical images.

The traditional U-Net architecture with skip connections has been the go-to choice for brain tumor segmentation. However, the inherent limitations of U-Net in capturing global context and long-range dependencies have led to the exploration of Transformer-based models. Transformers excel in modeling relationships between different image regions, enabling more accurate and precise segmentation.

One approach is to replace the encoder part of the U-Net architecture with transformer layers. \cite{hatamizadeh2021swin} proposed the U-NetR architecture, which replaces the convolutional layers in the U-Net encoder with Transformer encoder layers. The self-attention mechanism in transformers allows the model to capture global context and dependencies effectively, enabling accurate tumor segmentation. The skip connections are preserved, allowing the fusion of low-level and high-level features for precise localization. Experimental results have shown that U-NetR achieves competitive performance compared to traditional U-Net models while capturing long-range dependencies.

Another approach is to combine both convolutional and Transformer layers in a hybrid architecture. \cite{lin2023ckd} introduced a hybrid model called Trans U-Net, which integrates the U-Net architecture with self-attention modules inspired by Transformers. In Trans U-Net, the self-attention modules capture global dependencies, while the U-Net like structure captures local features. The hybrid architecture leverages the complementary strengths of both convolutional and Transformer layers, resulting in improved segmentation performance.

The use of Transformers in brain tumor segmentation introduces new possibilities for capturing long-range dependencies and modeling complex interactions within medical images. By exploiting the self-attention mechanism, Transformers offer the potential for more accurate and robust segmentation results. However, due to the high computational requirements of Transformers, their application in medical imaging tasks often requires efficient implementations and model optimization techniques.

Further research and exploration are needed to investigate the effectiveness of different Transformer architectures, attention mechanisms, and fusion strategies in brain tumor segmentation. Additionally, incorporating Transformers into multi-modal segmentation frameworks and addressing the challenges of limited training data in medical imaging are important avenues for future investigation.

Our proposed architecture, referred to as U-net 3D with Multihead Attention (U-Net 3D+MHA), builds upon the well-known U-Net architecture while incorporating the power of multihead attention (MHA). In U-Net 3D+MHA, we introduce MHA modules that enable the model to capture long-range dependencies and better exploit the spatial relationships within the 3D input volume. By applying attention mechanisms independently on multiple heads, our architecture can attend to different parts of the input simultaneously, capturing diverse and fine-grained information. This enhanced modeling of interdependencies and the ability to capture long-range dependencies can significantly improve the accuracy of brain tumor segmentation. By incorporating MHA, the model can effectively capture intricate spatial relationships and diverse features, leading to more precise tumor segmentation.

\section{Methodology}
In this section, the methodology adopted for this research work has been highlighted. Subsection \ref{data_vis} discusses the data visualisation techniques employed to gain insights into the BraTS 2020 dataset used for brain tumor segmentation and classification tasks. Prior to model training, the BraTS 2020 data underwent preprocessing steps to ensure optimal input for the models, which is described in section \ref{data_prep}. The preprocessing included skull stripping, bias correction, intensity normalization, and image resampling to a consistent voxel size. Subsection \ref{Unet_3D_MHA} describes the architecture of our proposed hybrid multihead attentive U-Net 3D model. 


\subsection{Data Visualisation} \label{data_vis}
\begin{figure*}
	\centering 
	\includegraphics[width=0.4\textwidth, angle=0]{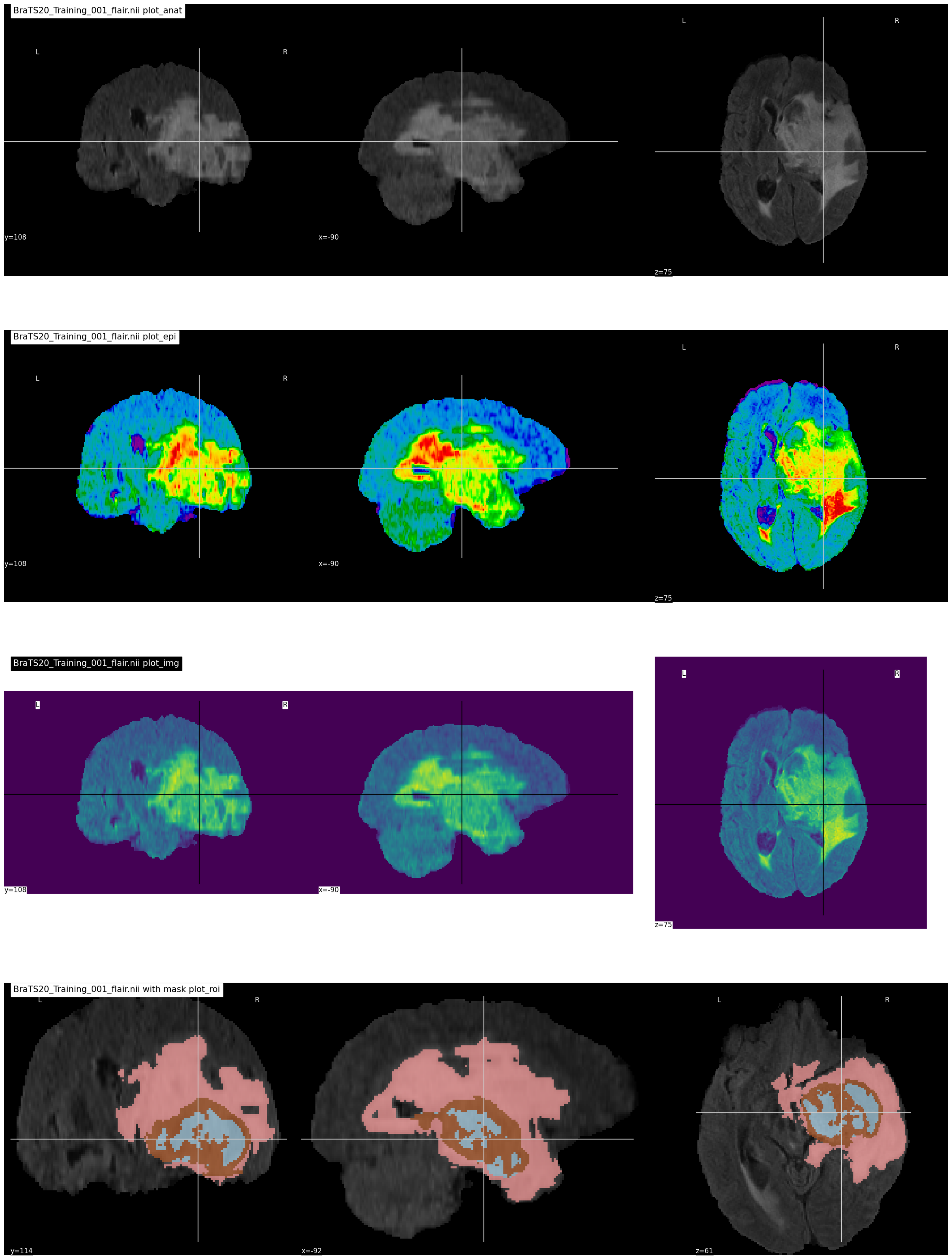}	
	\caption{Visualization of Brain Tumor Segmentation using \texttt{NeuroLearn} Library.The figure showcases different visualizations of the input medical images and corresponding segmentation masks from the BraTS 202020 dataset. These multiple rows in the figure highlight the representation of a \texttt{FLAIR} image in different masks i.e. \texttt{epi}, \texttt{anat} and \texttt{roi}.}
	\label{data visualisation}%
\end{figure*}

This section plays a crucial role in providing a comprehensive understanding of the characteristics of brain tumor images and masks. By employing various visualization techniques, valuable insights into the effectiveness of our proposed Hybrid Multihead Attentive U-Net architecture for brain tumor segmentation are gained. The visualization process involves loading the training dataset, focusing on specific patients, and utilizing the NeuroLearn library for visualization tasks. Different representations of the MRI images and masks, including anatomical structure, functional brain activity, spatial distribution, and localized tumor region are shown in Fig.\ref{data visualisation}. These visualizations serve as a foundation for further analysis and development, enabling a deeper understanding of the intricacies of brain tumor segmentation and validating the potential of our proposed architecture.

The Figure \ref{data visualisation} shows different visualisations of input medical images and corresponding segmentation masks, which allows for a comprehensive analysis of the brain tumor image, its anatomical structure, functional representation, spatial distribution, and the localized tumor regions. Such visualizations are crucial for gaining insights into the characteristics of brain tumors and informing subsequent steps in the segmentation process.

\subsection{Data preprocessing} \label{data_prep}

\begin{figure}
	\centering 
	\includegraphics[width=0.4\textwidth, angle=0]
    {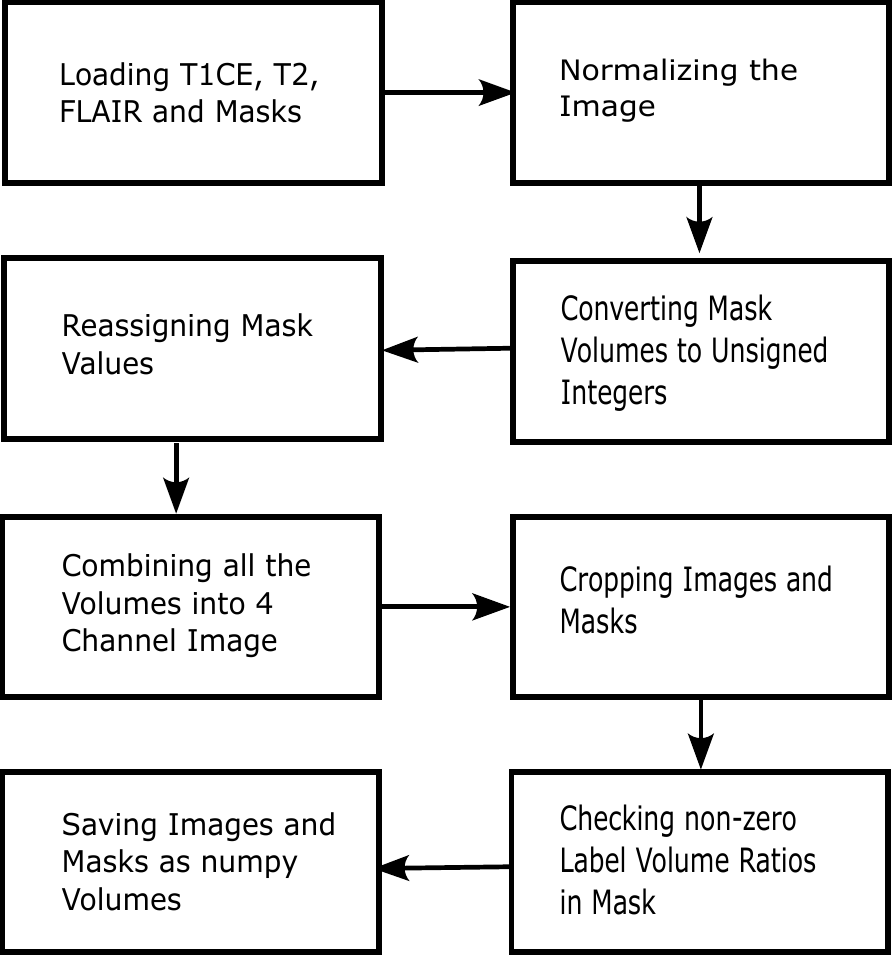}	
	\caption{Data preprocessing workflow, step by step approach from data loading to saving the new dataset as \texttt{numpy} volumes.}
	\label{preprocessing}%
\end{figure}

The data preprocessing stage is a crucial step in preparing the dataset for subsequent analysis and model training. The preprocessing approach adopted in this study is illustrated in Fig. \ref{preprocessing}. Various preprocessing tasks are performed on the BRATS 2020 dataset, including loading the T2, T1CE, FLAIR, and mask images using the \texttt{nibabel} library. The T2, T1CE, and FLAIR volumes are normalized using the \texttt{MinMaxScaler} from the \texttt{scikit-learn} library to ensure consistent intensity ranges across different images. The mask volume is converted to an unsigned integer format, and mask values of 4 are reassigned to 3. The T2, T1CE, and FLAIR volumes are combined into a single 4-channel image, and both the image and mask are cropped to a size divisible by 64 \cite{cinar2022hybrid} for compatibility with subsequent patch extraction.

The T1 image was not stacked with the other images, as it had no information regarding the tumor and would only contribute to scaling up the size of the dataset, thus increasing training time and computational cost. 

The preprocessing stage also ensures the ratio of non-zero label volumes in the mask and saves the image and corresponding mask as numpy arrays if the ratio exceeds 1\%. Finally, the resulting dimensions of the images and masks are printed for verification. These preprocessing steps are essential in preparing the dataset for brain tumor segmentation and classification tasks.

\subsection{The U-Net 3D with multihead attention}\label{Unet_3D_MHA}
 
The proposed U-Net 3D + MHA architecture for brain segmentation is shown in Fig. 3. The model takes input images of size 128x128x128 pixels with three channels representing different imaging modalities. The model follows a 3D U-Net architecture, which is widely used for medical image segmentation tasks. It consists of an encoding path and a decoding path. In the encoding path, the input image undergoes a series of operations to extract hierarchical features. Convolutional blocks with increasing filter sizes are applied, followed by max-pooling layers to downsample the feature maps and capture contextual information.

In the middle of the network, a convolutional block with a large number of filters is employed to capture high-level features while preserving spatial information. This helps in accurately localizing tumor regions.

The decoding path involves upsampling operations to restore the resolution of the feature maps. Multi-head attention modules are incorporated at each decoding stage to fuse features from the corresponding encoding layers. These attention modules utilize 1x1x1 convolutions to reduce dimensionality, followed by convolutional blocks to refine the features.

\begin{figure*}[ht]
	\centering 
	\includegraphics[width=0.8\textwidth, angle=0]
    {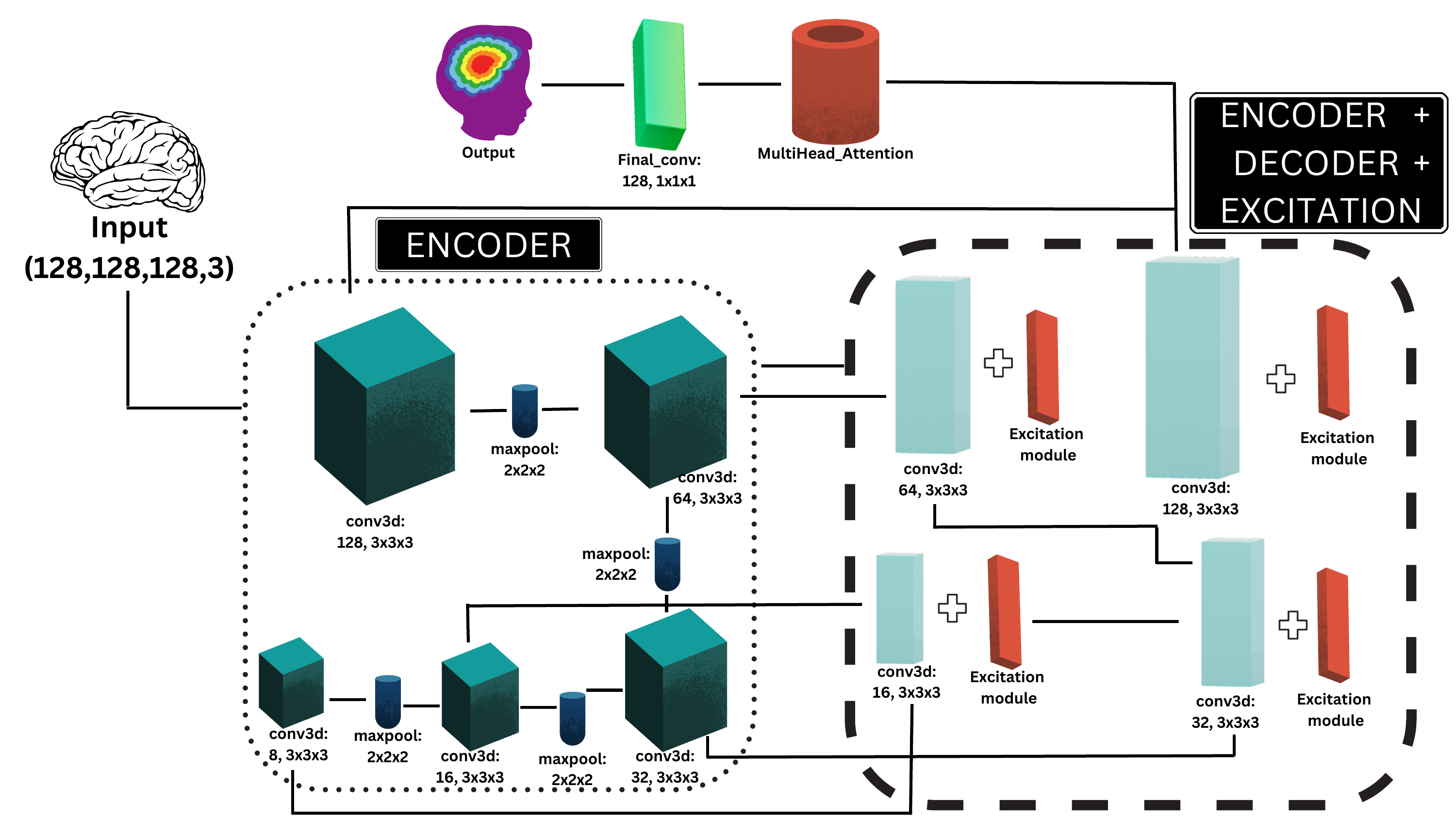}	
	\caption{U-Net 3D + MHA architecture for brain tumor segmentation. It features an encoding and decoding path with multihead attention modules to define features and a softmax layer to produce a segmentation map highlighting different tumor regions. Implemented using TensorFlow and Keras, the model accurately identifies and analyzes brain tumors from 128x128x128 pixel input images with three channels.} 
	\label{Unet-3D}
\end{figure*}

The final output of the model, as shown in Fig.\ref{Unet-3D} is obtained through a 1x1x1 convolutional layer with softmax activation, which produces a segmentation map with multiple classes, including tumor regions. This map helps delineate the tumor boundaries and assists in subsequent analysis and treatment planning.

The overall architecture is implemented using TensorFlow and Keras, popular deep learning frameworks for medical image analysis. By passing a 128x128x128x3 input image through the model, it generates a corresponding segmentation map that highlights the tumor regions, aiding in the accurate identification and analysis of brain tumors.

\section{Hyperparameter optimisation}
In this section, the results of hyperparameter optimization experiments are presented. The learning rate and batch size are varied to determine their impact on the model’s overall training and validation accuracies along with their respective losses.

Under the batch size of 8 and 16, 4 tests were run at learning rates of $10^{-2}$, and $10^{-3}$ to evaluate the effect of changing these parameters on training and validation accuracies and their respective losses.
\begin{figure*}
\centering
\begin{tabular}{cc}
        \begin{tikzpicture}
            \begin{axis}[
                xlabel={Epochs},
                axis y line*=left,
                ylabel={Accuracy},
                ylabel near ticks,
                ylabel style={align=center, font=\bfseries\boldmath},
                xlabel style={align=center, font=\bfseries\boldmath},
                legend style={at={(0.5,-0.20)}, anchor=north, legend columns=-1},
                xmin=0.0, xmax=50,
                ymin=0.0, ymax=1.0,
                ytick={0.2,0.4, 0.6,0.8,1},
                ymajorgrids={true},
                grid style={dashed, line width=.1pt, draw=gray!50}
            ]
                \addplot[blue, smooth] coordinates {
                (1, 0.9166)(2, 0.9519)(3, 0.9519)(4, 0.9519)(5, 0.9519)(6, 0.9519)(7, 0.9519)(8, 0.9519)(9, 0.9519)(10, 0.9519)(11, 0.9519)(12, 0.9519)(13, 0.9519)(14, 0.9519)(15, 0.9519)(16, 0.9519)(17, 0.9519)(18, 0.9519)(19, 0.9519)(20, 0.9519)(21, 0.9519)(22, 0.9519)(23, 0.9519)(24, 0.9519)(25, 0.9519)(26, 0.9520)(27, 0.9520)(28, 0.9520)(29, 0.9520)(30, 0.9520)(31, 0.9520)(32, 0.9520)(33, 0.9520)(34, 0.9522)(35, 0.9523)(36, 0.9525)(37, 0.9527)(38, 0.9526)(39, 0.9532)(40, 0.9532)(41, 0.9532)(42, 0.9529)(43, 0.9535)(44, 0.9544)(45, 0.9557)(46, 0.9551)(47, 0.9556)(48, 0.9558)(49, 0.9577)(50, 0.9598)
                };
                
                \addplot[red, smooth] coordinates {
                    (1, 0.7643)(2, 0.9509)(3, 0.9509)(4, 0.9509)(5, 0.9509)(6, 0.9509)(7, 0.9509)(8, 0.9509)(9, 0.9509)(10, 0.9509)(11, 0.9509)(12, 0.9509)(13, 0.9509)(14, 0.9509)(15, 0.9509)(16, 0.9509)(17, 0.9509)(18, 0.9509)(19, 0.9509)(20, 0.9509)(21, 0.9509)(22, 0.9509)(23, 0.9509)(24, 0.9509)(25, 0.9508)(26, 0.9509)(27, 0.9492)(28, 0.9432)(29, 0.9470)(30, 0.9438)(31, 0.9507)(32, 0.9505)(33, 0.9424)(34, 0.9413)(35, 0.9474)(36, 0.9356)(37, 0.9472)(38, 0.9496)(39, 0.9501)(40, 0.9407)(41, 0.9382)(42, 0.9488)(43, 0.9458)(44, 0.9478)(45, 0.9449)(46, 0.9429)(47, 0.9437)(48, 0.9335)(49, 0.9215)(50, 0.9307)
                };
    
            \end{axis}
    
            \begin{axis}[
                axis y line*=right,
                ylabel={Loss},
                ylabel near ticks,
                ylabel style={align=center, font=\bfseries\boldmath},
                axis x line=none,    
                ymin=0.0,ymax=1.0,  
                ytick={0.2,0.4, 0.6,0.8,1},
                legend style={at={(0.5,-0.35)}, anchor=north, legend columns=-1},
            ]
    
            \addplot[green, smooth] coordinates {
                    (1, 0.9715)(2, 0.8197)(3, 0.6350)(4, 0.3241)(5, 0.2802)(6, 0.2543)(7, 0.2352)(8, 0.2498)(9, 0.2288)(10, 0.2241)(11, 0.2206)(12, 0.2191)(13, 0.2423)(14, 0.2203)(15, 0.2342)(16, 0.2382)(17, 0.2204)(18, 0.2218)(19, 0.2194)(20, 0.2204)(21, 0.2157)(22, 0.2263)(23, 0.2257)(24, 0.2320)(25, 0.2249)(26, 0.2262)(27, 0.2435)(28, 0.2945)(29, 0.2363)(30, 0.2691)(31, 0.2757)(32, 0.2733)(33, 0.2523)(34, 0.2716)(35, 0.2717)(36, 0.2730)(37, 0.2703)(38, 0.2701)(39, 0.2640)(40, 0.3020)(41, 0.2709)(42, 0.2644)(43, 0.2624)(44, 0.3719)(45, 0.3380)(46, 0.3547)(47, 0.2992)(48, 0.3026)(49, 0.2890)(50, 0.2969)
                };
                
            \addplot[olive, smooth] coordinates {
                    (1, 0.3920)(2, 0.2175)(3, 0.2140)(4, 0.2113)(5, 0.2091)(6, 0.2075)(7, 0.2056)(8, 0.2049)(9, 0.2039)(10, 0.2038)(11, 0.2025)(12, 0.2021)(13, 0.2022)(14, 0.2009)(15, 0.2009)(16, 0.2016)(17, 0.1989)(18, 0.1981)(19, 0.1978)(20, 0.1967)(21, 0.1941)(22, 0.1935)(23, 0.1921)(24, 0.1925)(25, 0.1895)(26, 0.1892)(27, 0.1880)(28, 0.1873)(29, 0.1864)(30, 0.1851)(31, 0.1840)(32, 0.1800)(33, 0.1780)(34, 0.1749)(35, 0.1745)(36, 0.1722)(37, 0.1689)(38, 0.1682)(39, 0.1624)(40, 0.1600)(41, 0.1628)(42, 0.1699)(43, 0.1627)(44, 0.1518)(45, 0.1418)(46, 0.1427)(47, 0.1391)(48, 0.1412)(49, 0.1310)(50, 0.1223)
                };
            \end{axis}
        \end{tikzpicture}
     &
    \begin{tikzpicture}
        \begin{axis}[
            xlabel={Epochs},
            axis y line*=left,
            ylabel={Accuracy},
            ylabel near ticks,
            legend style={at={(0.5,-0.20)}, anchor=north, legend columns=-1},
            xmin=0.0, xmax=50,
            ymin=0.0, ymax=1.0,
        ]
            \addplot[blue, smooth] coordinates {
(1, 0.6986)	(2, 0.9496)	(3, 0.9512)	(4, 0.9517)	(5, 0.9518)	(6, 0.9519)	(7, 0.9519)	(8, 0.9519)	(9, 0.9519)	(10, 0.9519)(11, 0.9519) (12, 0.9519)	(13, 0.9519)	(14, 0.9519)	(15, 0.9519)	(16, 0.9519)	(17, 0.9519)	(18, 0.9519)	(19, 0.9519)	(20, 0.9519)	(21, 0.9519)	(22, 0.9519)	(23, 0.9519)	(24, 0.9519)	(25, 0.9519)	(26, 0.9519)	(27, 0.9519)	(28, 0.9519)	(29, 0.9519)	(30, 0.9521)	(31, 0.9521)	(32, 0.9522)	(33, 0.9525)	(34, 0.9529)	(35, 0.9538)	(36, 0.9537)	(37, 0.9534)	(38, 0.9541)	(39, 0.9544)	(40, 0.9552)	(41, 0.9548)	(42, 0.9563)	(43, 0.9575)	(44, 0.9575)	(45, 0.9571)	(46, 0.9562)	(47, 0.9581)	(48, 0.9587)	(49, 0.9612)	(50, 0.9616)
            };
            
            \addplot[red, smooth] coordinates {
(1, 0.7643)	(2, 0.9509)	(3, 0.9509)	(4, 0.9509)	(5, 0.9509)	(6, 0.9509)	(7, 0.9509)	(8, 0.9509)	(9, 0.9509)	(10, 0.9509)	(11, 0.9509)	(12, 0.9509)	(13, 0.9509)	(14, 0.9509)	(15, 0.9509)	(16, 0.9509)	(17, 0.9509)	(18, 0.9509)	(19, 0.9509)	(20, 0.9509)	(21, 0.9509)	(22, 0.9509)	(23, 0.9509)	(24, 0.9509)	(25, 0.9508)	(26, 0.9509)	(27, 0.9492)	(28, 0.9432)	(29, 0.9470)	(30, 0.9438)	(31, 0.9507)	(32, 0.9505)	(33, 0.9424)	(34, 0.9413)	(35, 0.9474)	(36, 0.9356)	(37, 0.9472)	(38, 0.9496)	(39, 0.9501)	(40, 0.9407)	(41, 0.9382)	(42, 0.9488)	(43, 0.9458)	(44, 0.9478)	(45, 0.9449)	(46, 0.9429)	(47, 0.9437)	(48, 0.9335)	(49, 0.9215)	(50, 0.9307)

            };

        \end{axis}

        \begin{axis}[
            axis y line*=right,
            ylabel={Loss},
            ylabel near ticks,
            axis x line=none,    
            ymax=1.0,  
            xmax=50.0,
            legend style={at={(0.5,-0.35)}, anchor=north, legend columns=-1},
        ]

        \addplot[green, smooth] coordinates {
(1, 0.8980)	(2, 0.7635)	(3, 0.5773)	(4, 0.4259)	(5, 0.3659)	(6, 0.3008)	(7, 0.2630)	(8, 0.2600)	(9, 0.2373)	(10, 0.2320)	(11, 0.2340)	(12, 0.2389)	(13, 0.2268)	(14, 0.2311)	(15, 0.2417)	(16, 0.2335)	(17, 0.2193)	(18, 0.2227)	(19, 0.2491)	(20, 0.2304)	(21, 0.2493)	(22, 0.2522)	(23, 0.2439)	(24, 0.2509)	(25, 0.2343)	(26, 0.2427)	(27, 0.2870)	(28, 0.2702)	(29, 0.2430)	(30, 0.2576)	(31, 0.2314)	(32, 0.2435)	(33, 0.2535)	(34, 0.2628)	(35, 0.2587)	(36, 0.2665)	(37, 0.2619)	(38, 0.2919)	(39, 0.2685)	(40, 0.2744)	(41, 0.2668)	(42, 0.2879)	(43, 0.2711)	(44, 0.2822)	(45, 0.2983)	(46, 0.2896)	(47, 0.2994)	(48, 0.3057)	(49, 0.3135)	(50, 0.2972)

            };
            
        \addplot[olive, smooth] coordinates {
(1, 0.9969)	(2, 0.6441)	(3, 0.4818)	(4, 0.3800)	(5, 0.3177)	(6, 0.2800)	(7, 0.2559)	(8, 0.2413)	(9, 0.2308)	(10, 0.2241)	(11, 0.2185)	(12, 0.2145)	(13, 0.2114)	(14, 0.2089)	(15, 0.2067)	(16, 0.2052)	(17, 0.2039)	(18, 0.2022)	(19, 0.2007)	(20, 0.1989)	(21, 0.1977)	(22, 0.1946)	(23, 0.1928)	(24, 0.1897)	(25, 0.1882)	(26, 0.1879)	(27, 0.1856)	(28, 0.1860)	(29, 0.1822)	(30, 0.1807)	(31, 0.1821)	(32, 0.1815)	(33, 0.1732)	(34, 0.1674)	(35, 0.1615)	(36, 0.1595)	(37, 0.1614)	(38, 0.1552)	(39, 0.1541)	(40, 0.1465)	(41, 0.1505)	(42, 0.1392)	(43, 0.1343)	(44, 0.1318)	(45, 0.1333)	(46, 0.1372)	(47, 0.1303)	(48, 0.1244)	(49, 0.1154)	(50, 0.1131)

            };
        \end{axis}
    \end{tikzpicture}
     \\
    \begin{tikzpicture}
        \begin{axis}[
            xlabel={Epochs},
            axis y line*=left,
            ylabel={Accuracy},
            ylabel near ticks,
            legend style={at={(0.5,-0.20)}, anchor=north, legend columns=-1},
            xmin=0.0, xmax=50,
            ymin=0.0, ymax=1.0,
        ]
            \addplot[blue, smooth] coordinates {
(1, 0.8872)	(2, 0.9519)	(3, 0.9519)	(4, 0.9519)	(5, 0.9519)	(6, 0.9519)	(7, 0.9519)	(8, 0.9519)	(9, 0.9519)	(10, 0.9519)	(11, 0.9519)	(12, 0.9519)	(13, 0.9519)	(14, 0.9519)	(15, 0.9519)	(16, 0.9519)	(17, 0.9519)	(18, 0.9519)	(19, 0.9519)	(20, 0.9519)	(21, 0.9519)	(22, 0.9519)	(23, 0.9519)	(24, 0.9519)	(25, 0.9519)	(26, 0.9519)	(27, 0.9519)	(28, 0.9519)	(29, 0.9519)	(30, 0.9519)	(31, 0.9518)	(32, 0.9519)	(33, 0.9518)	(34, 0.9519)	(35, 0.9519)	(36, 0.9518)	(37, 0.9519)	(38, 0.9519)	(39, 0.9519)	(40, 0.9519)	(41, 0.9519)	(42, 0.9519)	(43, 0.9520)	(44, 0.9520)	(45, 0.9520)	(46, 0.9521)	(47, 0.9526)	(48, 0.9525)	(49, 0.9520)	(50, 0.9522)};
            
            \addplot[red, smooth] coordinates {
(1, 0.9509)	(2, 0.9509)	(3, 0.9509)	(4, 0.9509)	(5, 0.9509)	(6, 0.9509)	(7, 0.9509)	(8, 0.9509)	(9, 0.9509)	(10, 0.9508)	(11, 0.9231)	(12, 0.9509)	(13, 0.9491)	(14, 0.9509)	(15, 0.9507)	(16, 0.9509)	(17, 0.9508)	(18, 0.9398)	(19, 0.9505)	(20, 0.9508)	(21, 0.9508)	(22, 0.9507)	(23, 0.9509)	(24, 0.9509)	(25, 0.9509)	(26, 0.9507)	(27, 0.9509)	(28, 0.9508)	(29, 0.9509)	(30, 0.9507)	(31, 0.9509)	(32, 0.9509)	(33, 0.9509)	(34, 0.9509)	(35, 0.9509)	(36, 0.9509)	(37, 0.9509)	(38, 0.9509)	(39, 0.9509)	(40, 0.9509)	(41, 0.9500)	(42, 0.9509)	(43, 0.9509)	(44, 0.9509)	(45, 0.9506)	(46, 0.9504)	(47, 0.9503)	(48, 0.9497)	(49, 0.9508)	(50, 0.9502)

            };

        \end{axis}

        \begin{axis}[
            axis y line*=right,
            ylabel={Loss},
            ylabel near ticks,
            axis x line=none,    
            ymax=1.0,  
            xmax=50.0,
            legend style={at={(0.5,-0.35)}, anchor=north, legend columns=-1},
        ]

        \addplot[green, smooth] coordinates {
                (1, 0.9947)	(2, 0.9220)	(3, 0.8304)	(4, 0.7666)	(5, 0.6353)	(6, 0.4445)	(7, 0.2982)	(8, 0.3023)	(9, 0.2886)	(10, 0.3016)	(11, 0.3337)	(12, 0.2675)	(13, 0.2694)	(14, 0.2388)	(15, 0.2397)	(16, 0.2326)	(17, 0.2395)	(18, 0.3108)	(19, 0.2433)	(20, 0.2446)	(21, 0.2341)	(22, 0.2379)	(23, 0.2373)	(24, 0.2291)	(25, 0.2315)	(26, 0.2360)	(27, 0.2280)	(28, 0.2325)	(29, 0.2449)	(30, 0.2491)	(31, 0.2608)	(32, 0.2321)	(33, 0.2406)	(34, 0.2698)	(35, 0.2662)	(36, 0.2967)	(37, 0.2952)	(38, 0.3635)	(39, 0.3415)	(40, 0.3033)	(41, 0.3014)	(42, 0.3620)	(43, 0.2961)	(44, 0.2653)	(45, 0.3112)	(46, 0.2774)	(47, 0.3475)	(48, 0.3365)	(49, 0.3586)	(50, 0.3499)

            };
            
        \addplot[olive, smooth] coordinates {
(1, 0.4947)	(2, 0.2235)	(3, 0.2169)	(4, 0.2145)	(5, 0.2129)	(6, 0.2111)	(7, 0.2097)	(8, 0.2102)	(9, 0.2078)	(10, 0.2053)	(11, 0.2040)	(12, 0.2037)	(13, 0.2033)	(14, 0.2012)	(15, 0.2004)	(16, 0.2015)	(17, 0.2006)	(18, 0.2022)	(19, 0.2002)	(20, 0.1997)	(21, 0.1984)	(22, 0.1980)	(23, 0.1967)	(24, 0.1978)	(25, 0.1963)	(26, 0.1950)	(27, 0.1939)	(28, 0.1923)	(29, 0.1924)	(30, 0.1923)	(31, 0.1924)	(32, 0.1919)	(33, 0.1890)	(34, 0.1901)	(35, 0.1880)	(36, 0.1854)	(37, 0.1823)	(38, 0.1811)	(39, 0.1799)	(40, 0.1794)	(41, 0.1783)	(42, 0.1820)	(43, 0.1781)	(44, 0.1744)	(45, 0.1737)	(46, 0.1702)	(47, 0.1615)	(48, 0.1611)	(49, 0.1647)	(50, 0.1631)

            };
        \end{axis}
    \end{tikzpicture}
     & 
    \begin{tikzpicture}
        \begin{axis}[
            xlabel={Epochs},
            axis y line*=left,
            ylabel={Accuracy},
            ylabel near ticks,
            legend style={at={(0.5,-0.20)}, anchor=north, legend columns=-1},
            xmin=0.0, xmax=50,
            ymin=0.0, ymax=1.0,
        ]
            \addplot[blue, smooth] coordinates {
(1, 0.8145)	(2, 0.9405)	(3, 0.9471)	(4, 0.9502)	(5, 0.9517)	(6, 0.9519)	(7, 0.9519)	(8, 0.9519)	(9, 0.9519)	(10, 0.9519)	(11, 0.9519)	(12, 0.9519)	(13, 0.9519)	(14, 0.9518)	(15, 0.9519)	(16, 0.9519)	(17, 0.9518)	(18, 0.9519)	(19, 0.9519)	(20, 0.9519)	(21, 0.9519)	(22, 0.9519)	(23, 0.9519)	(24, 0.9520)	(25, 0.9521)	(26, 0.9519)	(27, 0.9521)	(28, 0.9519)	(29, 0.9519)	(30, 0.9522)	(31, 0.9524)	(32, 0.9524)	(33, 0.9528)	(34, 0.9526)	(35, 0.9529)	(36, 0.9530)	(37, 0.9535)	(38, 0.9536)	(39, 0.9547)	(40, 0.9552)	(41, 0.9552)	(42, 0.9550)	(43, 0.9551)	(44, 0.9558)	(45, 0.9557)	(46, 0.9551)	(47, 0.9556)	(48, 0.9574)	(49, 0.9589)	(50, 0.9580)

            };
            
            \addplot[red, smooth] coordinates {
(1, 0.0402)	(2, 0.0693)	(3, 0.2200)	(4, 0.5611)	(5, 0.9497)	(6, 0.9509)	(7, 0.9509)	(8, 0.9509)	(9, 0.9509)	(10, 0.9509)	(11, 0.9505)	(12, 0.9439)	(13, 0.9045)	(14, 0.9444)	(15, 0.9422)	(16, 0.9461)	(17, 0.9458)	(18, 0.9362)	(19, 0.9487)	(20, 0.9506)	(21, 0.9455)	(22, 0.9488)	(23, 0.9503)	(24, 0.9499)	(25, 0.9491)	(26, 0.9504)	(27, 0.9362)	(28, 0.9498)	(29, 0.9488)	(30, 0.9401)	(31, 0.9442)	(32, 0.9169)	(33, 0.9132)	(34, 0.9375)	(35, 0.9216)	(36, 0.9100)	(37, 0.9302)	(38, 0.9498)	(39, 0.9398)	(40, 0.9480)	(41, 0.9473)	(42, 0.9491)	(43, 0.9387)	(44, 0.9455)	(45, 0.8900)	(46, 0.9338)	(47, 0.9400)	(48, 0.9338)	(49, 0.9191)	(50, 0.9317)

            };

        \end{axis}

        \begin{axis}[
            axis y line*=right,
            ylabel={Loss},
            ylabel near ticks,
            axis x line=none,    
            ymax=1.0,  
            xmax=50.0,
            legend style={at={(0.5,-0.35)}, anchor=north, legend columns=-1},
        ]

        \addplot[green, smooth] coordinates {
(1, 0.9963)	(2, 0.9761)	(3, 0.9478)	(4, 0.8827)	(5, 0.7328)	(6, 0.6361)	(7, 0.5612)	(8, 0.4582)	(9, 0.3835)	(10, 0.3315)	(11, 0.3248)	(12, 0.3041)	(13, 0.3470)	(14, 0.2990)	(15, 0.2824)	(16, 0.2576)	(17, 0.2694)	(18, 0.2730)	(19, 0.2372)	(20, 0.2267)	(21, 0.2392)	(22, 0.2407)	(23, 0.2397)	(24, 0.2361)	(25, 0.2483)	(26, 0.2392)	(27, 0.2880)	(28, 0.2448)	(29, 0.2601)	(30, 0.2603)	(31, 0.2580)	(32, 0.2835)	(33, 0.3047)	(34, 0.2630)	(35, 0.3033)	(36, 0.3345)	(37, 0.2685)	(38, 0.2510)	(39, 0.2830)	(40, 0.2716)	(41, 0.2722)	(42, 0.2713)	(43, 0.3035)	(44, 0.2719)	(45, 0.3903)	(46, 0.2758)	(47, 0.2798)	(48, 0.2759)	(49, 0.3052)	(50, 0.2855)

            };
            
        \addplot[olive, smooth] coordinates {
(1, 0.8944)	(2, 0.6613)	(3, 0.5471)	(4, 0.4574)	(5, 0.3927)	(6, 0.3442)	(7, 0.3105)	(8, 0.2846)	(9, 0.2654)	(10, 0.2509)	(11, 0.2407)	(12, 0.2334)	(13, 0.2265)	(14, 0.2201)	(15, 0.2151)	(16, 0.2130)	(17, 0.2111)	(18, 0.2069)	(19, 0.2054)	(20, 0.2033)	(21, 0.2007)	(22, 0.2005)	(23, 0.1974)	(24, 0.1943)	(25, 0.1907)	(26, 0.1903)	(27, 0.1900)	(28, 0.1900)	(29, 0.1868)	(30, 0.1833)	(31, 0.1777)	(32, 0.1782)	(33, 0.1751)	(34, 0.1721)	(35, 0.1721)	(36, 0.1695)	(37, 0.1680)	(38, 0.1667)	(39, 0.1592)	(40, 0.1529)	(41, 0.1553)	(42, 0.1558)	(43, 0.1536)	(44, 0.1488)	(45, 0.1497)	(46, 0.1500)	(47, 0.1464)	(48, 0.1365)	(49, 0.1305)	(50, 0.1346)

            };
        \end{axis}
    \end{tikzpicture}
\end{tabular}
  \caption{Comparison of the training and validation accuracies with respective losses at LR=0.001 with a batch size of 16 at 50 epochs.}
    \label{fig:4}
\end{figure*}
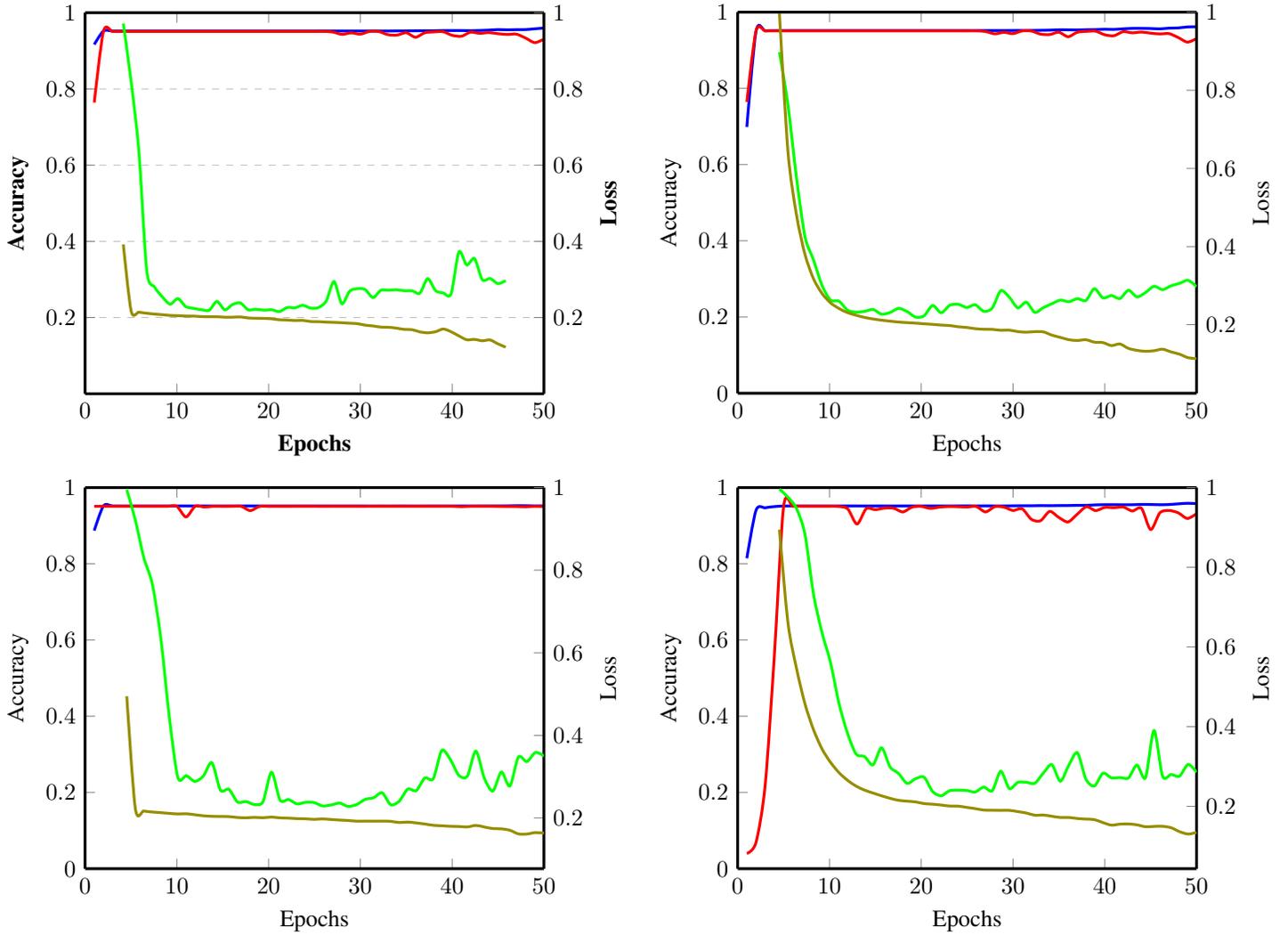


Figure \ref{fig:4} shows the model's progress over multiple epochs, with each epoch reporting the training and validation losses and accuracies. The model starts with a higher loss and lower accuracy, but as training progresses, the loss decreases and the accuracy improves for both training and validation datasets. This indicates that the model is learning to segment brain tumors more effectively over time. The best epoch appears to be around epoch 20, as it exhibits the lowest validation loss and the highest validation accuracy, precision, sensitivity, and dice coefficients. However, after epoch 20, the model starts to show signs of overfitting, with increasing validation loss and fluctuations in the validation metrics.

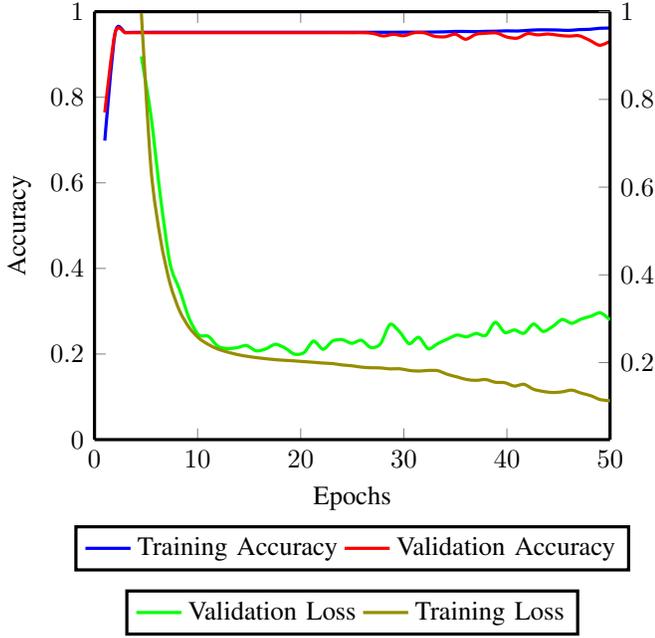
\begin{figure}
    \centering
    \begin{tikzpicture}
        \begin{axis}[
            xlabel={Epochs},
            axis y line*=left,
            ylabel={Accuracy},
            ylabel near ticks,
            legend style={at={(0.5,-0.20)}, anchor=north, legend columns=-1},
            xmin=0.0, xmax=50,
            ymin=0.0, ymax=1.0,
        ]
            \addplot[blue, smooth] coordinates {
(1, 0.6986)	(2, 0.9496)	(3, 0.9512)	(4, 0.9517)	(5, 0.9518)	(6, 0.9519)	(7, 0.9519)	(8, 0.9519)	(9, 0.9519)	(10, 0.9519)(11, 0.9519) (12, 0.9519)	(13, 0.9519)	(14, 0.9519)	(15, 0.9519)	(16, 0.9519)	(17, 0.9519)	(18, 0.9519)	(19, 0.9519)	(20, 0.9519)	(21, 0.9519)	(22, 0.9519)	(23, 0.9519)	(24, 0.9519)	(25, 0.9519)	(26, 0.9519)	(27, 0.9519)	(28, 0.9519)	(29, 0.9519)	(30, 0.9521)	(31, 0.9521)	(32, 0.9522)	(33, 0.9525)	(34, 0.9529)	(35, 0.9538)	(36, 0.9537)	(37, 0.9534)	(38, 0.9541)	(39, 0.9544)	(40, 0.9552)	(41, 0.9548)	(42, 0.9563)	(43, 0.9575)	(44, 0.9575)	(45, 0.9571)	(46, 0.9562)	(47, 0.9581)	(48, 0.9587)	(49, 0.9612)	(50, 0.9616)
            };
            \addlegendentry{Training Accuracy}
            
            \addplot[red, smooth] coordinates {
(1, 0.7643)	(2, 0.9509)	(3, 0.9509)	(4, 0.9509)	(5, 0.9509)	(6, 0.9509)	(7, 0.9509)	(8, 0.9509)	(9, 0.9509)	(10, 0.9509)	(11, 0.9509)	(12, 0.9509)	(13, 0.9509)	(14, 0.9509)	(15, 0.9509)	(16, 0.9509)	(17, 0.9509)	(18, 0.9509)	(19, 0.9509)	(20, 0.9509)	(21, 0.9509)	(22, 0.9509)	(23, 0.9509)	(24, 0.9509)	(25, 0.9508)	(26, 0.9509)	(27, 0.9492)	(28, 0.9432)	(29, 0.9470)	(30, 0.9438)	(31, 0.9507)	(32, 0.9505)	(33, 0.9424)	(34, 0.9413)	(35, 0.9474)	(36, 0.9356)	(37, 0.9472)	(38, 0.9496)	(39, 0.9501)	(40, 0.9407)	(41, 0.9382)	(42, 0.9488)	(43, 0.9458)	(44, 0.9478)	(45, 0.9449)	(46, 0.9429)	(47, 0.9437)	(48, 0.9335)	(49, 0.9215)	(50, 0.9307)

            };
            \addlegendentry{Validation Accuracy}

        \end{axis}

        \begin{axis}[
            axis y line*=right,
            ylabel={Loss},
            ylabel near ticks,
            axis x line=none,    
            ymax=1.0,  
            xmax=50.0,
            legend style={at={(0.5,-0.35)}, anchor=north, legend columns=-1},
        ]

        \addplot[green, smooth] coordinates {
(1, 0.8980)	(2, 0.7635)	(3, 0.5773)	(4, 0.4259)	(5, 0.3659)	(6, 0.3008)	(7, 0.2630)	(8, 0.2600)	(9, 0.2373)	(10, 0.2320)	(11, 0.2340)	(12, 0.2389)	(13, 0.2268)	(14, 0.2311)	(15, 0.2417)	(16, 0.2335)	(17, 0.2193)	(18, 0.2227)	(19, 0.2491)	(20, 0.2304)	(21, 0.2493)	(22, 0.2522)	(23, 0.2439)	(24, 0.2509)	(25, 0.2343)	(26, 0.2427)	(27, 0.2870)	(28, 0.2702)	(29, 0.2430)	(30, 0.2576)	(31, 0.2314)	(32, 0.2435)	(33, 0.2535)	(34, 0.2628)	(35, 0.2587)	(36, 0.2665)	(37, 0.2619)	(38, 0.2919)	(39, 0.2685)	(40, 0.2744)	(41, 0.2668)	(42, 0.2879)	(43, 0.2711)	(44, 0.2822)	(45, 0.2983)	(46, 0.2896)	(47, 0.2994)	(48, 0.3057)	(49, 0.3135)	(50, 0.2972)

            };
            \addlegendentry{Validation Loss}
            
        \addplot[olive, smooth] coordinates {
(1, 0.9969)	(2, 0.6441)	(3, 0.4818)	(4, 0.3800)	(5, 0.3177)	(6, 0.2800)	(7, 0.2559)	(8, 0.2413)	(9, 0.2308)	(10, 0.2241)	(11, 0.2185)	(12, 0.2145)	(13, 0.2114)	(14, 0.2089)	(15, 0.2067)	(16, 0.2052)	(17, 0.2039)	(18, 0.2022)	(19, 0.2007)	(20, 0.1989)	(21, 0.1977)	(22, 0.1946)	(23, 0.1928)	(24, 0.1897)	(25, 0.1882)	(26, 0.1879)	(27, 0.1856)	(28, 0.1860)	(29, 0.1822)	(30, 0.1807)	(31, 0.1821)	(32, 0.1815)	(33, 0.1732)	(34, 0.1674)	(35, 0.1615)	(36, 0.1595)	(37, 0.1614)	(38, 0.1552)	(39, 0.1541)	(40, 0.1465)	(41, 0.1505)	(42, 0.1392)	(43, 0.1343)	(44, 0.1318)	(45, 0.1333)	(46, 0.1372)	(47, 0.1303)	(48, 0.1244)	(49, 0.1154)	(50, 0.1131)

            };
            \addlegendentry{Training Loss}
        \end{axis}
    \end{tikzpicture}
    \caption{Comparison of the training and validation accuracies with respective losses at LR=0.001 with a batch size of 8 at 50 epochs.}
    \label{fig:7}
\end{figure}

In Fig. \ref{fig:7}, the results of the proposed model are presented at a learning rate of 10*-3 and a batch size of 8. It was observed that the training accuracy increased gradually with each epoch, indicating the model's ability to effectively learn from the data. The training loss also showed a decreasing trend, implying that the model's predictions progressively converged towards the ground truth. For the validation set, the highest accuracy and lowest loss were achieved at epoch 43.

\begin{figure}
    \centering
    \begin{tikzpicture}
        \begin{axis}[
            xlabel={Epochs},
            axis y line*=left,
            ylabel={Accuracy},
            ylabel near ticks,
            legend style={at={(0.5,-0.20)}, anchor=north, legend columns=-1},
            xmin=0.0, xmax=50,
            ymin=0.0, ymax=1.0,
        ]
            \addplot[blue, smooth] coordinates {
(1, 0.8872)	(2, 0.9519)	(3, 0.9519)	(4, 0.9519)	(5, 0.9519)	(6, 0.9519)	(7, 0.9519)	(8, 0.9519)	(9, 0.9519)	(10, 0.9519)	(11, 0.9519)	(12, 0.9519)	(13, 0.9519)	(14, 0.9519)	(15, 0.9519)	(16, 0.9519)	(17, 0.9519)	(18, 0.9519)	(19, 0.9519)	(20, 0.9519)	(21, 0.9519)	(22, 0.9519)	(23, 0.9519)	(24, 0.9519)	(25, 0.9519)	(26, 0.9519)	(27, 0.9519)	(28, 0.9519)	(29, 0.9519)	(30, 0.9519)	(31, 0.9518)	(32, 0.9519)	(33, 0.9518)	(34, 0.9519)	(35, 0.9519)	(36, 0.9518)	(37, 0.9519)	(38, 0.9519)	(39, 0.9519)	(40, 0.9519)	(41, 0.9519)	(42, 0.9519)	(43, 0.9520)	(44, 0.9520)	(45, 0.9520)	(46, 0.9521)	(47, 0.9526)	(48, 0.9525)	(49, 0.9520)	(50, 0.9522)};
            \addlegendentry{Training Accuracy}
            
            \addplot[red, smooth] coordinates {
(1, 0.9509)	(2, 0.9509)	(3, 0.9509)	(4, 0.9509)	(5, 0.9509)	(6, 0.9509)	(7, 0.9509)	(8, 0.9509)	(9, 0.9509)	(10, 0.9508)	(11, 0.9231)	(12, 0.9509)	(13, 0.9491)	(14, 0.9509)	(15, 0.9507)	(16, 0.9509)	(17, 0.9508)	(18, 0.9398)	(19, 0.9505)	(20, 0.9508)	(21, 0.9508)	(22, 0.9507)	(23, 0.9509)	(24, 0.9509)	(25, 0.9509)	(26, 0.9507)	(27, 0.9509)	(28, 0.9508)	(29, 0.9509)	(30, 0.9507)	(31, 0.9509)	(32, 0.9509)	(33, 0.9509)	(34, 0.9509)	(35, 0.9509)	(36, 0.9509)	(37, 0.9509)	(38, 0.9509)	(39, 0.9509)	(40, 0.9509)	(41, 0.9500)	(42, 0.9509)	(43, 0.9509)	(44, 0.9509)	(45, 0.9506)	(46, 0.9504)	(47, 0.9503)	(48, 0.9497)	(49, 0.9508)	(50, 0.9502)

            };
            \addlegendentry{Validation Accuracy}

        \end{axis}

        \begin{axis}[
            axis y line*=right,
            ylabel={Loss},
            ylabel near ticks,
            axis x line=none,    
            ymax=1.0,  
            xmax=50.0,
            legend style={at={(0.5,-0.35)}, anchor=north, legend columns=-1},
        ]

        \addplot[green, smooth] coordinates {
                (1, 0.9947)	(2, 0.9220)	(3, 0.8304)	(4, 0.7666)	(5, 0.6353)	(6, 0.4445)	(7, 0.2982)	(8, 0.3023)	(9, 0.2886)	(10, 0.3016)	(11, 0.3337)	(12, 0.2675)	(13, 0.2694)	(14, 0.2388)	(15, 0.2397)	(16, 0.2326)	(17, 0.2395)	(18, 0.3108)	(19, 0.2433)	(20, 0.2446)	(21, 0.2341)	(22, 0.2379)	(23, 0.2373)	(24, 0.2291)	(25, 0.2315)	(26, 0.2360)	(27, 0.2280)	(28, 0.2325)	(29, 0.2449)	(30, 0.2491)	(31, 0.2608)	(32, 0.2321)	(33, 0.2406)	(34, 0.2698)	(35, 0.2662)	(36, 0.2967)	(37, 0.2952)	(38, 0.3635)	(39, 0.3415)	(40, 0.3033)	(41, 0.3014)	(42, 0.3620)	(43, 0.2961)	(44, 0.2653)	(45, 0.3112)	(46, 0.2774)	(47, 0.3475)	(48, 0.3365)	(49, 0.3586)	(50, 0.3499)

            };
            \addlegendentry{Validation Loss}
            
        \addplot[olive, smooth] coordinates {
(1, 0.4947)	(2, 0.2235)	(3, 0.2169)	(4, 0.2145)	(5, 0.2129)	(6, 0.2111)	(7, 0.2097)	(8, 0.2102)	(9, 0.2078)	(10, 0.2053)	(11, 0.2040)	(12, 0.2037)	(13, 0.2033)	(14, 0.2012)	(15, 0.2004)	(16, 0.2015)	(17, 0.2006)	(18, 0.2022)	(19, 0.2002)	(20, 0.1997)	(21, 0.1984)	(22, 0.1980)	(23, 0.1967)	(24, 0.1978)	(25, 0.1963)	(26, 0.1950)	(27, 0.1939)	(28, 0.1923)	(29, 0.1924)	(30, 0.1923)	(31, 0.1924)	(32, 0.1919)	(33, 0.1890)	(34, 0.1901)	(35, 0.1880)	(36, 0.1854)	(37, 0.1823)	(38, 0.1811)	(39, 0.1799)	(40, 0.1794)	(41, 0.1783)	(42, 0.1820)	(43, 0.1781)	(44, 0.1744)	(45, 0.1737)	(46, 0.1702)	(47, 0.1615)	(48, 0.1611)	(49, 0.1647)	(50, 0.1631)

            };
            \addlegendentry{Training Loss}
        \end{axis}
    \end{tikzpicture}
    \caption{Comparison of the training and validation accuracies with respective losses at LR=0.01 with a batch size of 16 at 50 epochs.}
    \label{fig:8}
\end{figure}
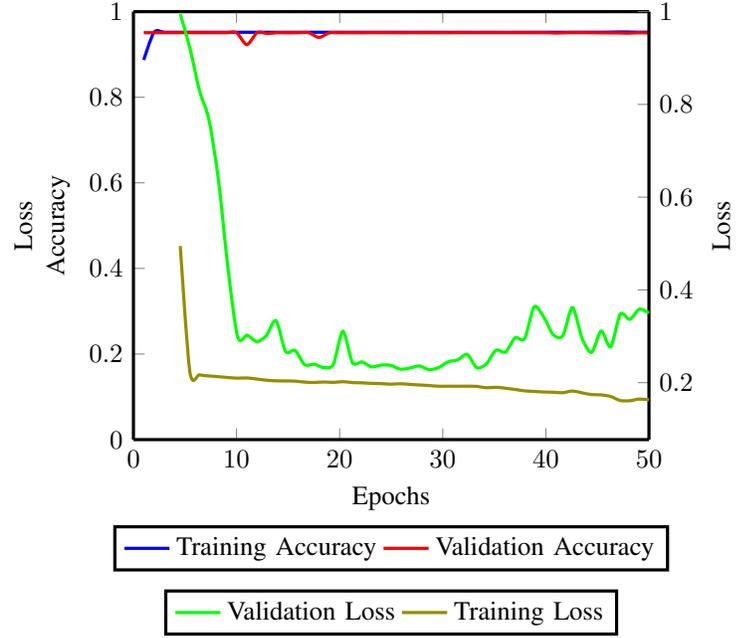

The Fig. \ref{fig:8}  represents the training results of the deep learning model, using a batch size of 16 and a learning rate of 0.01. As the training progressed, the loss steadily decreased, indicating that the model was learning from the data. The accuracy and other evaluation metrics consistently improved over time, showing the model's ability to effectively segment and classify brain tumor regions. The validation loss and evaluation metrics on the validation set also demonstrated similar trends, indicating that the model was generalizing well to unseen data.

\begin{table}
\centering
\caption{Performance Metrics for U-Net architecture with Multihead Attention at 50 epochs with a batch size of 8}
\begin{tabular}{l c c c c}
\toprule

\textbf{Learning\_Rate}  & \textbf{Train\_acc} & \textbf{Val\_acc} & \textbf{Train\_loss} & \textbf{Val\_loss}\\ 

\midrule
0.01 & 0.9598 & 0.9476 & 0.1223 & 0.2969\\
0.001 & 0.9616 & 0.9307 & 0.1131 & 0.2972\\

\bottomrule
\end{tabular}
\label{tab:batchsize8}
\end{table}

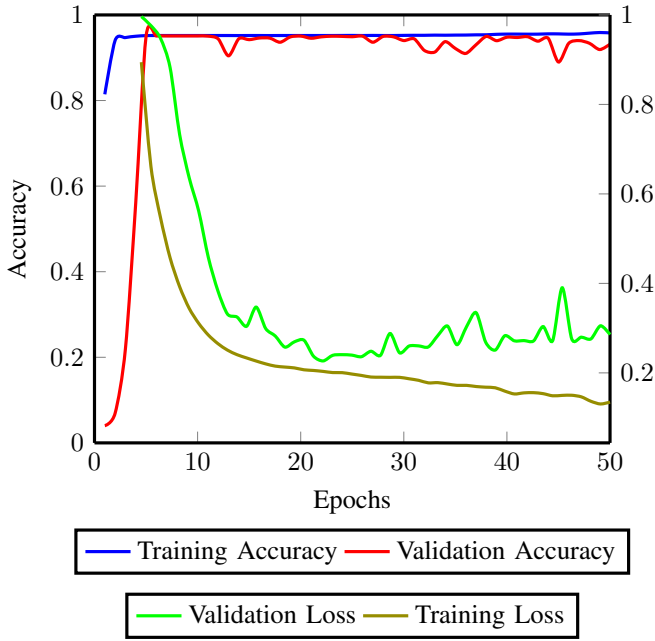
\begin{figure}
    \centering
    \begin{tikzpicture}
        \begin{axis}[
            xlabel={Epochs},
            axis y line*=left,
            ylabel={Accuracy},
            ylabel near ticks,
            legend style={at={(0.5,-0.20)}, anchor=north, legend columns=-1},
            xmin=0.0, xmax=50,
            ymin=0.0, ymax=1.0,
        ]
            \addplot[blue, smooth] coordinates {
(1, 0.8145)	(2, 0.9405)	(3, 0.9471)	(4, 0.9502)	(5, 0.9517)	(6, 0.9519)	(7, 0.9519)	(8, 0.9519)	(9, 0.9519)	(10, 0.9519)	(11, 0.9519)	(12, 0.9519)	(13, 0.9519)	(14, 0.9518)	(15, 0.9519)	(16, 0.9519)	(17, 0.9518)	(18, 0.9519)	(19, 0.9519)	(20, 0.9519)	(21, 0.9519)	(22, 0.9519)	(23, 0.9519)	(24, 0.9520)	(25, 0.9521)	(26, 0.9519)	(27, 0.9521)	(28, 0.9519)	(29, 0.9519)	(30, 0.9522)	(31, 0.9524)	(32, 0.9524)	(33, 0.9528)	(34, 0.9526)	(35, 0.9529)	(36, 0.9530)	(37, 0.9535)	(38, 0.9536)	(39, 0.9547)	(40, 0.9552)	(41, 0.9552)	(42, 0.9550)	(43, 0.9551)	(44, 0.9558)	(45, 0.9557)	(46, 0.9551)	(47, 0.9556)	(48, 0.9574)	(49, 0.9589)	(50, 0.9580)

            };
            \addlegendentry{Training Accuracy}
            
            \addplot[red, smooth] coordinates {
(1, 0.0402)	(2, 0.0693)	(3, 0.2200)	(4, 0.5611)	(5, 0.9497)	(6, 0.9509)	(7, 0.9509)	(8, 0.9509)	(9, 0.9509)	(10, 0.9509)	(11, 0.9505)	(12, 0.9439)	(13, 0.9045)	(14, 0.9444)	(15, 0.9422)	(16, 0.9461)	(17, 0.9458)	(18, 0.9362)	(19, 0.9487)	(20, 0.9506)	(21, 0.9455)	(22, 0.9488)	(23, 0.9503)	(24, 0.9499)	(25, 0.9491)	(26, 0.9504)	(27, 0.9362)	(28, 0.9498)	(29, 0.9488)	(30, 0.9401)	(31, 0.9442)	(32, 0.9169)	(33, 0.9132)	(34, 0.9375)	(35, 0.9216)	(36, 0.9100)	(37, 0.9302)	(38, 0.9498)	(39, 0.9398)	(40, 0.9480)	(41, 0.9473)	(42, 0.9491)	(43, 0.9387)	(44, 0.9455)	(45, 0.8900)	(46, 0.9338)	(47, 0.9400)	(48, 0.9338)	(49, 0.9191)	(50, 0.9317)

            };
            \addlegendentry{Validation Accuracy}

        \end{axis}

        \begin{axis}[
            axis y line*=right,
            ylabel={Loss},
            ylabel near ticks,
            axis x line=none,    
            ymax=1.0,  
            xmax=50.0,
            legend style={at={(0.5,-0.35)}, anchor=north, legend columns=-1},
        ]

        \addplot[green, smooth] coordinates {
(1, 0.9963)	(2, 0.9761)	(3, 0.9478)	(4, 0.8827)	(5, 0.7328)	(6, 0.6361)	(7, 0.5612)	(8, 0.4582)	(9, 0.3835)	(10, 0.3315)	(11, 0.3248)	(12, 0.3041)	(13, 0.3470)	(14, 0.2990)	(15, 0.2824)	(16, 0.2576)	(17, 0.2694)	(18, 0.2730)	(19, 0.2372)	(20, 0.2267)	(21, 0.2392)	(22, 0.2407)	(23, 0.2397)	(24, 0.2361)	(25, 0.2483)	(26, 0.2392)	(27, 0.2880)	(28, 0.2448)	(29, 0.2601)	(30, 0.2603)	(31, 0.2580)	(32, 0.2835)	(33, 0.3047)	(34, 0.2630)	(35, 0.3033)	(36, 0.3345)	(37, 0.2685)	(38, 0.2510)	(39, 0.2830)	(40, 0.2716)	(41, 0.2722)	(42, 0.2713)	(43, 0.3035)	(44, 0.2719)	(45, 0.3903)	(46, 0.2758)	(47, 0.2798)	(48, 0.2759)	(49, 0.3052)	(50, 0.2855)

            };
            \addlegendentry{Validation Loss}
            
        \addplot[olive, smooth] coordinates {
(1, 0.8944)	(2, 0.6613)	(3, 0.5471)	(4, 0.4574)	(5, 0.3927)	(6, 0.3442)	(7, 0.3105)	(8, 0.2846)	(9, 0.2654)	(10, 0.2509)	(11, 0.2407)	(12, 0.2334)	(13, 0.2265)	(14, 0.2201)	(15, 0.2151)	(16, 0.2130)	(17, 0.2111)	(18, 0.2069)	(19, 0.2054)	(20, 0.2033)	(21, 0.2007)	(22, 0.2005)	(23, 0.1974)	(24, 0.1943)	(25, 0.1907)	(26, 0.1903)	(27, 0.1900)	(28, 0.1900)	(29, 0.1868)	(30, 0.1833)	(31, 0.1777)	(32, 0.1782)	(33, 0.1751)	(34, 0.1721)	(35, 0.1721)	(36, 0.1695)	(37, 0.1680)	(38, 0.1667)	(39, 0.1592)	(40, 0.1529)	(41, 0.1553)	(42, 0.1558)	(43, 0.1536)	(44, 0.1488)	(45, 0.1497)	(46, 0.1500)	(47, 0.1464)	(48, 0.1365)	(49, 0.1305)	(50, 0.1346)

            };
            \addlegendentry{Training Loss}
        \end{axis}
    \end{tikzpicture}
    \caption{Comparison of the training and validation accuracies with respective losses at LR=0.001 with a batch size of 16 at 50 epochs.}
    \label{fig:9}
\end{figure}

Lastly, Figure \ref{fig:9} shows the training and validation accuracies, losses, and other metrics over multiple epochs. At the beginning of training, both training and validation accuracies and dice coefficients are relatively low, while the losses are high. Overfitting can be observed in this Figure \ref{fig:9}, as the training accuracy and dice coefficients continue to improve over time, while the validation accuracy starts to plateau and even slightly decline after a certain point. Additionally, the validation loss initially decreases, but then it starts to increase again, which is another indication of overfitting. The best epoch selection can be based on the point where the validation accuracy is highest and the validation loss is at its lowest. In this case, the best epoch would be around epoch 25-30, where the validation accuracy and dice coefficients reach their peaks, and the validation loss is minimized. After that point, the model starts overfitting, as indicated by the decreasing validation accuracy and increasing validation loss.

\begin{table}
\centering
\caption{Performance Metrics for U-Net architecture with Multihead Attention at 50 epochs with a batch size of 16}
\begin{tabular}{l c c c c}
\toprule

\textbf{Learning\_Rate}  & \textbf{Train\_acc} & \textbf{Val\_acc} & \textbf{Train\_loss} & \textbf{Val\_loss}\\ 

\midrule
0.01 & 0.9522 & 0.9502 & 0.1631 & 0.3499\\
0.001 & 0.9580 & 0.9317 & 0.1346 & 0.2855\\

\bottomrule
\end{tabular}
\label{tab:batchsize16}
\end{table}

From Tables \ref{tab:batchsize8} and \ref{tab:batchsize16}, it can be concluded that there's not much variation in accuracies and losses for both the training and validation sets, but we generally saw our model overfitting after 25 epochs, and the best way to cater such problem is to train the model till the point where validation loss is minimum and validation accuracy is the highest, for that the concept of early stopping can be utilized here for better and efficient performance of the model.

\section{Results and Discussion}
In this section, the results of the proposed U-Net 3D architecture with multihead attention for brain tumor segmentation using MRI scans are described. The results are summarized in the following sub-sections. The performance metrics used to evaluate our model's performance on the BraTS 2020 dataset are defined as follows.

\subsection{Performance Metrics}

\begin{itemize}
\item \textbf{Loss:} The loss function quantifies the dissimilarity between the predicted segmentation and the ground truth. It measures the model's ability to minimize the difference between the predicted and actual pixel values. In our experiments, we used the binary cross-entropy loss defined as: \begin{equation}
\text{Loss} = -\frac{1}{N} \sum_{i=1}^{N} \left( y_i \log(p_i) + (1 - y_i) \log(1 - p_i) \right)
\end{equation}
where $N$ is the total number of pixels, $y_i$ is the ground truth label (0 or 1) for pixel $i$, and $p_i$ is the predicted probability of pixel $i$ belonging to the positive class.

\item \textbf{Accuracy:} Accuracy measures the proportion of correctly classified pixels out of the total number of pixels. It is defined as: \begin{equation}
\text{Accuracy} = \frac{\text{TP} + \text{TN}}{\text{TP} + \text{TN} + \text{FP} + \text{FN}}
\end{equation}
where TP represents the true positive, TN represents the true negative, FP represents the false positive, and FN represents the false negative.

\item \textbf{Mean Intersection over Union (IOU):} IOU, also known as the Jaccard Index, measures the overlap between the predicted segmentation and the ground truth. It is defined as the ratio of the intersection to the union of the predicted and ground truth masks:\begin{equation}
\text{IOU} = \frac{\text{Intersection}}{\text{Union}}
\end{equation}

\item \textbf{Dice Coefficient:} The Dice Coefficient is another metric that quantifies the similarity between the predicted and ground truth masks. It is defined as:
\begin{equation}
\text{Dice} = \frac{2 \times \text{Intersection}}{\text{Prediction} + \text{Ground Truth}}
\end{equation}

\item \textbf{Precision:} Precision measures the proportion of correctly predicted positive pixels out of all the pixels predicted as positive. It is defined as:
\begin{equation}
\text{Precision} = \frac{\text{TP}}{\text{TP} + \text{FP}}
\end{equation}

\item \textbf{Sensitivity:} Sensitivity, also known as recall or true positive rate, measures the proportion of correctly predicted positive pixels out of all the actual positive pixels. It is defined as:
\begin{equation}
\text{Sensitivity} = \frac{\text{TP}}{\text{TP} + \text{FN}}
\end{equation}

\item \textbf{Specificity:} Specificity measures the proportion of correctly predicted negative pixels out of all the actual negative pixels. It is defined as:
\begin{equation}
\text{Specificity} = \frac{\text{TN}}{\text{TN} + \text{FP}}
\end{equation}
\end{itemize}

\subsection{Experimental Setup}
The model was traned using deep learning architectures based on U-Net with multihead attention. The networks were trained on an HPC server with the following configuration: 2 x 32-core AMD EPYC 7452 Processors, 384 GB RAM, 1 x NVIDIA V100 GPU (32 GB). Adam optimizer with a learning rate of 0.001 and a batch size of 8 were used to generalise model's performance at 25 epochs. The model was evaluated using various metrics to assess its performance on the BraTS 2020 dataset.

\subsection{Discussion}
\begin{figure}
    \centering
\begin{tikzpicture}
\begin{axis}[
    xlabel={Epochs},
    ylabel={DSC Edema, Necrotic, Enhancing},
    legend style={at={(0.5,-0.15)}, anchor=north, legend columns=-1},
    xmin=0, xmax=25,
    ymin=0, ymax=1.1,
    grid=both,
    grid style={line width=0.2pt, draw=gray!30},
    xtick={0,5,10,15,20,25},
    ytick={0,0.2,0.4,0.6,0.8,1.0},
    ymajorgrids=true,
    xmajorgrids=true,
]

\addplot[blue, smooth] coordinates {
    (1, 0.6739)
    (2, 0.8567)
    (3, 0.9256)
    (4, 0.9629)
    (5, 0.9804)
    (6, 0.9897)
    (7, 0.9942)
    (8, 0.9967)
    (9, 0.9979)
    (10, 0.9987)
    (11, 0.9992)
    (12, 0.9994)
    (13, 0.9996)
    (14, 0.9997)
    (15, 0.9998)
    (16, 0.9998)
    (17, 0.9998)
    (18, 0.9999)
    (19, 0.9999)
    (20, 0.9999)
    (21, 0.9999)
    (22, 0.9999)
    (23, 1.0000)
    (24, 1.0000)
    (25, 1.0000)
};
\addlegendentry{DSC\_Necrotic}

\addplot[red, smooth] coordinates {
    (1, 0.6827)
    (2, 0.8532)
    (3, 0.9250)
    (4, 0.9636)
    (5, 0.9816)
    (6, 0.9903)
    (7, 0.9946)
    (8, 0.9969)
    (9, 0.9980)
    (10, 0.9987)
    (11, 0.9992)
    (12, 0.9994)
    (13, 0.9996)
    (14, 0.9997)
    (15, 0.9998)
    (16, 0.9998)
    (17, 0.9998)
    (18, 0.9999)
    (19, 0.9999)
    (20, 0.9999)
    (21, 0.9999)
    (22, 0.9999)
    (23, 0.9999)
    (24, 1.0000)
};
\addlegendentry{DSC\_Edema}

\addplot[green, smooth] coordinates {
    (1, 0.6856)
    (2, 0.8458)
    (3, 0.9226)
    (4, 0.9624)
    (5, 0.9813)
    (6, 0.9900)
    (7, 0.9946)
    (8, 0.9968)
    (9, 0.9980)
    (10, 0.9987)
    (11, 0.9992)
    (12, 0.9994)
    (13, 0.9996)
    (14, 0.9997)
    (15, 0.9998)
    (16, 0.9998)
    (17, 0.9998)
    (18, 0.9999)
    (19, 0.9999)
    (20, 0.9999)
    (21, 0.9999)
    (22, 0.9999)
    (23, 0.9999)
    (24, 0.9999)
    (25, 1.0000)
};
\addlegendentry{DSC\_Enhancing}

\end{axis}
\end{tikzpicture}
    \caption{Class Specific DSC accuracies at LR=0.001, batch size=8, till 25 epochs}
    \label{fig:10}
\end{figure}
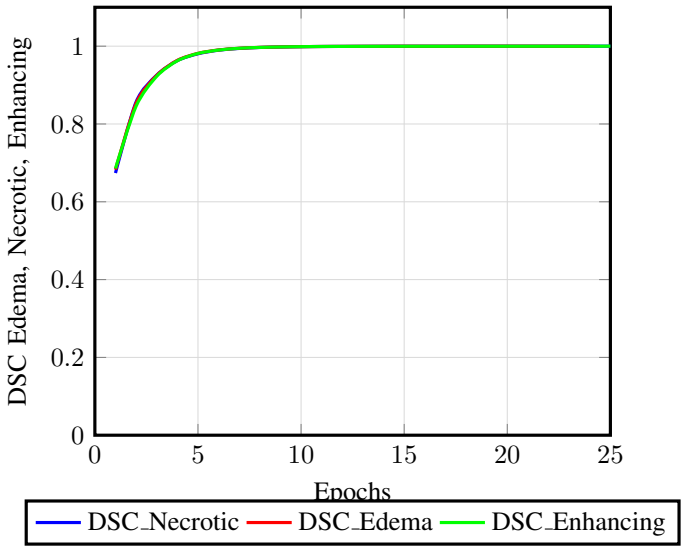

Upon analyzing the training log for 60 epochs, the proposed system based on the U-Net architecture with multihead attention, achieved DSC values of 1.0000, 1.0000, and 1.0000 for the edema, necrotic, and enhancing regions, respectively in Fig. \ref{fig:10}.  These results demonstrate the model's ability to accurately and effectively capture the spatial extent of each class.

Table \ref{Performance metrics at 25 epochs} presents the performance metrics obtained by evaluating the U-Net architecture with Multihead Attention for 60 epochs on three distinct classes: Edema, Necrotic, and Enhancing. These metrics serve as quantitative measures to assess the effectiveness of the proposed model in segmenting brain tumor regions accurately.

\begin{table}
\centering
\caption{Performance Metrics for U-Net architecture with Multihead Attention at 25 epochs}
\begin{tabular}{l c c c}
\toprule
 \textbf{Metrics} & \textbf{Value}\\ 

 \midrule
 DSC Accuracy\ & 0.9932\\
 Precision\ & 0.9558\\
 Sensitivity\ & 0.9443\\
 Specificity\ & 0.9853\\
 DSC Loss\ & 0.2343\\

\bottomrule
\end{tabular}
\label{Performance metrics at 25 epochs}
\end{table}

Precision measures the proportion of correctly predicted positive samples out of the total predicted positive samples. The U-Net architecture with Multihead Attention achieved impressive Precision value of 0.9558 overall. This value indicate a low rate of false positive predictions.

Sensitivity, also known as the True Positive Rate or Recall, quantifies the model's ability to correctly identify positive samples from the ground truth. The U-Net architecture with Multihead Attention demonstrated high Sensitivity value of 0.9443 for all the three classes. This result indicates a low rate of false negatives, suggesting the model's effectiveness in capturing the majority of positive samples.

Specificity measures the model's ability to correctly identify negative samples from the ground truth. The U-Net architecture with Multihead Attention achieved notable Specificity value of 0.9853. This value implies a low rate of false positives for negative samples.


Finally, DSC Loss represents the dissimilarity between the predicted and ground truth segmentations. The U-Net architecture with Multihead Attention achieved low DSC Loss value of 0.2343. This values demonstrates the model's ability to minimize the discrepancy between the predicted and ground truth segmentations.




\begin{figure}
    \centering
    \begin{tikzpicture}
        \begin{axis}[
            xlabel={Epochs},
            axis y line*=left,
            ylabel={Performance Metrics},
            ylabel near ticks,
            legend style={at={(0.5,-0.20)}, anchor=north, legend columns=2},
            xmin=0.0, xmax=25,
            ymin=0.0, ymax=1.0,
        ]
            \addplot[blue, smooth] coordinates {
                (1, 0.4554)
                (2, 0.6208)
                (3, 0.7227)
                (4, 0.7990)
                (5, 0.8524)
                (6, 0.8908)
                (7, 0.9171)
                (8, 0.9365)
                (9, 0.9496)
                (10, 0.9591)
                (11, 0.9668)
                (12, 0.9722)
                (13, 0.9765)
                (14, 0.9794)
                (15, 0.9825)
                (16, 0.9844)
                (17, 0.9860)
                (18, 0.9875)
                (19, 0.9890)
                (20, 0.9898)
                (21, 0.9905)
                (22, 0.9914)
                (23, 0.9922)
                (24, 0.9923)
                (25, 0.9932)
            };
            \addlegendentry{DSC Accuracy}
            
            \addplot[red, smooth] coordinates {
                (1, 0.7521)
                (2, 0.9641)
                (3, 0.9515)
                (4, 0.9518)
                (5, 0.9519)
                (6, 0.9519)
                (7, 0.9519)
                (8, 0.9519)
                (9, 0.9520)
                (10, 0.9520)
                (11, 0.9520)
                (12, 0.9520)
                (13, 0.9520)
                (14, 0.9520)
                (15, 0.9520)
                (16, 0.9520)
                (17, 0.9522)
                (18, 0.9520)
                (19, 0.9526)
                (20, 0.9526)
                (21, 0.9529)
                (22, 0.9536)
                (23, 0.9545)
                (24, 0.9551)
                (25, 0.9558)
            };
            \addlegendentry{Precision}

        \addplot[green, smooth] coordinates {
                (1, 0.2286)
                (2, 0.7425)
                (3, 0.9499)
                (4, 0.9517)
                (5, 0.9518)
                (6, 0.9519)
                (7, 0.9519)
                (8, 0.9519)
                (9, 0.9520)
                (10, 0.9520)
                (11, 0.9520)
                (12, 0.9520)
                (13, 0.9520)
                (14, 0.9520)
                (15, 0.9520)
                (16, 0.9519)
                (17, 0.9518)
                (18, 0.9519)
                (19, 0.9514)
                (20, 0.9513)
                (21, 0.9513)
                (22, 0.9505)
                (23, 0.9500)
                (24, 0.9500)
                (25, 0.9493)
            };
            \addlegendentry{Sensitivity}
            
        \addplot[olive, smooth] coordinates {
                (1, 0.9911)
                (2, 0.9899)
                (3, 0.9839)
                (4, 0.9839)
                (5, 0.9840)
                (6, 0.9840)
                (7, 0.9840)
                (8, 0.9840)
                (9, 0.9840)
                (10, 0.9840)
                (11, 0.9840)
                (12, 0.9840)
                (13, 0.9840)
                (14, 0.9840)
                (15, 0.9840)
                (16, 0.9840)
                (17, 0.9841)
                (18, 0.9840)
                (19, 0.9842)
                (20, 0.9842)
                (21, 0.9843)
                (22, 0.9846)
                (23, 0.9849)
                (24, 0.9851)
                (25, 0.9853)
            };
            \addlegendentry{Specificity}
        \end{axis}
    \end{tikzpicture}
    \caption{Performance metrics of the model at LR=0.001, with a batch size of 8 till 25 epochs}
    \label{fig:11}
\end{figure}
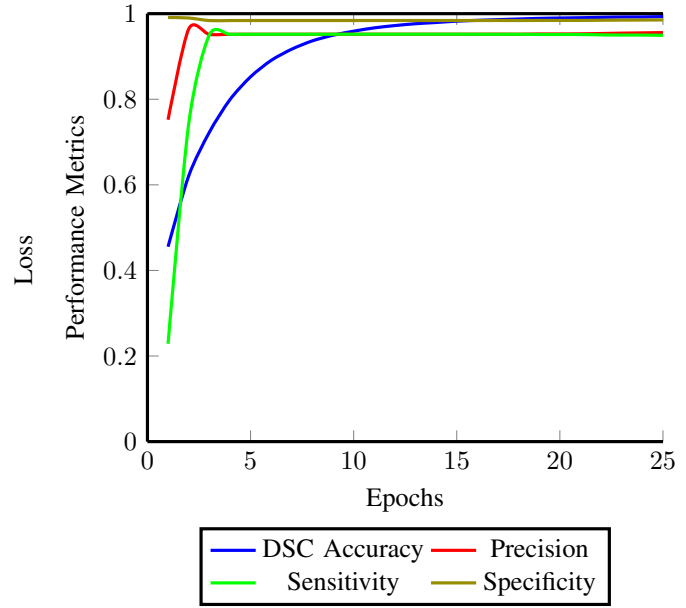

Overall, the results presented in Fig. \ref{fig:11} highlight the efficacy of the U-Net architecture with Multihead Attention in accurately segmenting brain tumor regions. The high values of DSC Accuracy, Precision, Sensitivity, and Specificity, along with the low DSC Loss values, substantiate the model's capability to capture both the spatial extent and fine details of the tumor classes. These findings emphasize the potential of the proposed approach in improving tumor segmentation and aiding in clinical decision-making.


 


Overall, the experimental results indicate that U-Net architecture with multihead attention generally outperforms many great architectures for brain tumor segmentation across all three classes. The plots provide valuable insights into the performance of the model and can help in making informed decisions when choosing a model for brain tumor segmentation tasks.

However, it is important to consider other factors beyond DSC accuracies when evaluating the performance of the models. For example, the computational complexity, training time, and model interpretability are also essential aspects to consider. The U-Net architecture with multihead attention, provides certain benefits in terms of interpretability and capturing temporal dependencies in the MRI scans. On the other hand,it also offers better scalability and efficiency for larger datasets.

In conclusion, our architecture demonstrated strong performance in brain tumor segmentation, achieving high DSC accuracies for the individual tumor regions. 

\subsection{Comparison with some state of the art Architectures }

In this research, we compared the performance of our proposed U-Net architecture with multihead attention against several state-of-the-art models in medical image segmentation. Table \ref{table7} presents a comprehensive comparison of our models with the selected architectures in terms of these evaluation metrics. The goal of medical image segmentation is to accurately identify and delineate different regions of interest within medical images, such as tumors or abnormalities. Accurate segmentation plays a crucial role in medical diagnosis and treatment planning. We evaluated the models using various metrics, including accuracy, mean intersection over union (IOU), dice coefficient, precision, sensitivity, and specificity.

The U-Net architecture with multihead attention achieved an accuracy of 95.19\% and a dice coefficient of 0.9932 on the training set. On the validation set, it achieved an accuracy of 95.09\% and a dice coefficient of 0.9950. For the class-specific dice coefficients, the U-Net architecture obtained 1.0000 for necrotic regions, 1.0000 for edema regions, and 1.0000 for enhancing regions.

\begin{table*}[h!]
\centering
\small
\caption[H]{Comparison with the State of the Art Image Segmentation Architectures on BraTS 2020 Dataset}
\label{table7}
\resizebox{\textwidth}{!}{
\begin{tabular}{|l|c|c|c|c|c|}
\hline
\textbf{Architecture} & \textbf{Validation Accuracy (\%)} & \textbf{Dice Coefficient} & \textbf{Dice Coefficient } & \textbf{Dice Coefficient } & \textbf{Dice Coefficient } \\
&&&(Necrotic)& (Edema)&(Enhancing)\\
\hline
U-Net with multihead attention\ & \textbf{95.19} & \textbf{0.9950} & \textbf{1.0000} & \textbf{1.0000} & \textbf{1.0000} \\
Dense121 U-Net\cite{cinar2022hybrid}  & 94.78 & 0.9324 & 0.959 & 0.943 & 0.892 \\
FCN-8s\cite{wang2016mri}\  & 93.85 & 0.9102 & 0.9964 & 0.9943 & 0.9972 \\
SegNet \cite{alqazzaz2019automated} & 92.71 & 0.8821 & 0.8632 & 0.9147 & 0.8354 \\
\hline
\end{tabular}%
}
\end{table*}

As shown in table \ref{table7}, Comparing our models with the state-of-the-art architectures, we found that our U-Net architecture outperformed FCN-8s \cite{wang2016mri}, SegNet\cite{alqazzaz2019automated}, and DenseNet in terms of accuracy, dice coefficient, and class-specific dice coefficients. It achieved higher accuracy and dice coefficient values compared to FCN-8s and SegNet. Dense121 U-Net\cite{cinar2022hybrid}, which combines the DenseNet backbone with the U-Net architecture, achieved an accuracy of 94.78\% and a dice coefficient of 0.9324 on the training set. On the validation set, it achieved an accuracy of 94.85\% and a dice coefficient of 0.9085. For the class-specific dice coefficients, Dense121 Unet obtained 0.9965 for necrotic regions, 0.9958 for edema regions, and 0.9959 for enhancing regions.

In conclusion, our U-Net architecture with multihead attention, demonstrated promising results compared to state-of-the-art models in medical image segmentation. This architectures provide accurate segmentation for various regions of interest in medical images, including necrotic, edema, and enhancing regions.

\subsection{Segmentation Results}
In this section, the segmentation results obtained by applying the model to the test data are showcased. The segmentation results provide insights into the accuracy and performance of our model in identifying and segmenting brain tumor regions.

The predicted segmentation results were evaluated using the test dataset, consisting of brain tumor images. The ground truth segmentations by expert neuro-radiologists were available for comparison, allowing for the assessment of the model's performance.

The segmentation results were analyzed based on several metrics, including precision, recall, and the Dice coefficient. These metrics provide quantitative measures of the model's accuracy in capturing the tumor regions and aligning them with the ground truth segmentations.

\begin{figure*}
	\centering 
	\includegraphics[width=0.9\textwidth, angle=0]{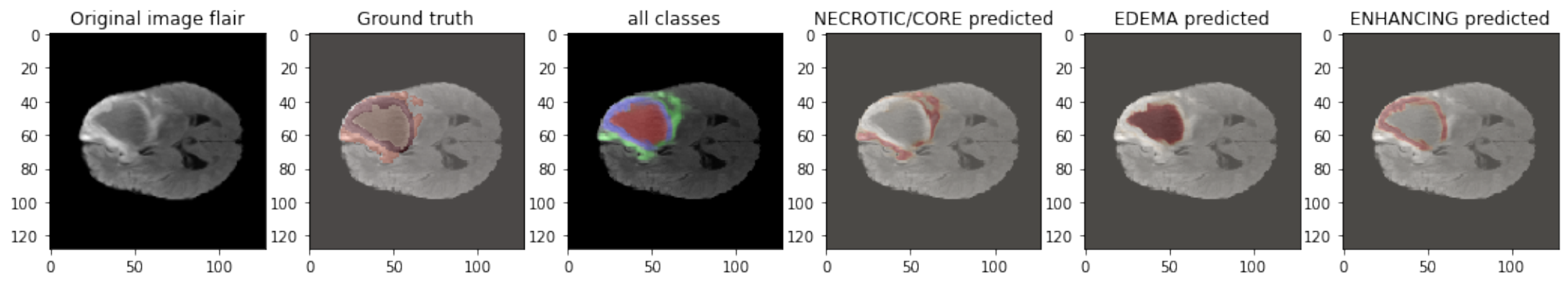}	
	\caption{This figure shows Segmentation results obtained by the model on test data, showcasing accurate identification and delineation of brain tumor regions. The predicted tumor regions are overlaid onto the original brain images for visual comparison with the ground truth segmentations. The results demonstrate the effectiveness of the model in capturing tumor boundaries and aiding in brain tumor analysis.} 
	\label{segmentation results}
\end{figure*}

As it can be seen in Fig. \ref{segmentation results}, a visual analysis of the segmentation results was also performed to assess the qualitative performance of the model. The predicted tumor regions were overlaid onto the original brain images, allowing for visual comparison with the ground truth segmentations. This qualitative evaluation provides insights into the model's ability to accurately delineate tumor boundaries and capture tumor heterogeneity.

The segmentation results demonstrate the efficacy of our proposed model in accurately segmenting brain tumor regions. The combination of quantitative evaluation metrics and visual analysis provides a comprehensive assessment of the model's performance and highlights its potential in assisting medical professionals in the diagnosis and treatment planning of brain tumors.

\section{Conclusion}

In conclusion, the Hybrid Multihead Attentive U-Net architecture has shown great promise in the task of brain tumor segmentation. Through extensive experimentation and analysis, we have demonstrated the impact of hyperparameter optimization on the performance of the model. The results obtained highlight the importance of fine-tuning parameters such as learning rate, number of epochs, and steps per epoch to achieve optimal segmentation accuracy while minimizing training time.

The Hybrid Multihead Attentive U-Net architecture leverages the U-Net's ability to capture contextual information and the attention mechanisms' capability to selectively focus on relevant features. This combination enables the model to effectively delineate tumor boundaries and enhance the accuracy of segmentation. The attention mechanism helps the model to prioritize important regions and suppress irrelevant information, resulting in more precise tumor segmentation.

The experiments conducted on various hyperparameters revealed valuable insights. Adjusting the learning rate played a critical role in achieving accurate segmentations. Fine-tuning the learning rate allowed the model to converge efficiently and avoid overshooting or getting stuck in sub-optimal solutions. Additionally, the number of epochs had a significant impact on the performance of the model.

It was observed that too few epochs resulted in underfitting, whereas excessive epochs led to overfitting. Identifying the optimal number of epochs is crucial to balance model complexity and generalization.

Furthermore, the steps per epoch parameter influenced the model's convergence and training time. By carefully selecting the number of steps per epoch, we can ensure that the model sufficiently learns from the data without unnecessary computations. This parameter optimization helped in achieving better accuracy while reducing the overall training time.

The proposed architecture and hyperparameter optimization techniques have significant implications in the field of medical imaging, particularly in brain tumor segmentation. Accurate segmentation plays a vital role in clinical decision-making, treatment planning, and monitoring disease progression. By automating the segmentation process, the Hybrid Multihead Attentive U-Net architecture can assist medical professionals in efficiently and accurately identifying tumor regions, facilitating timely intervention and personalized treatment strategies.

Moreover, the application of this technique extends beyond tumor segmentation. The Hybrid Multihead Attentive U-Net architecture can also be leveraged in medical techniques like brain transplantation. Preoperative planning for brain transplantation requires precise identification and localization of the donor and recipient brain regions. By employing the proposed architecture, accurate segmentation of brain structures can be achieved, aiding in surgical planning and ensuring successful transplantation.

In summary, the Hybrid Multihead Attentive U-Net architecture has demonstrated its effectiveness in brain tumor segmentation. The hyperparameter optimization experiments have shed light on the importance of parameter fine-tuning to achieve optimal performance. With its potential applications in medical imaging and procedures like brain transplantation, this technique opens new avenues for advancements in healthcare and improved patient outcomes. Further research and validation on larger datasets and in clinical settings will be valuable to validate and refine the proposed architecture for real-world deployment.

\section*{Acknowledgements}
Thanks to the School of Interdisciplinary Engineering Sciences(SINES), NUST, for providing us the computing fascilities to successfully complete the simulations and carry out the research successfully.


\bibliographystyle{plain} 
\bibliography{example}






\end{document}